\documentclass{article}
\usepackage{epsfig}

\date{\today}

\usepackage{amsfonts}
\usepackage{amsmath}

\numberwithin{equation}{section}

\usepackage{epsfig}
\usepackage{amsmath}
\usepackage{amsfonts}\tolerance=10000
\usepackage{graphicx}

\newcommand{\insertplot}[5]{\begin{figure}
 \hfill\hbox to 0.05in{\vbox to #5in{\vfill 
 \inputplot{#1}{#4}{#5}}\hfill}
 \hfill\vspace{-.1in}
 \caption{#2}\label{#3}
 \end{figure}}
 \newcommand{\inputplot}[3]{
 \special{ps: plotfile #1}
}
\newcounter{fig}   
\newcommand{\lp}{\left(}
\def\ga{\gamma}
\newcommand{\rp}{\right)}

\newcommand{\beq}{\begin{equation}}
\newcommand{\eeq}{\end{equation}}
\newcommand{\beqs}{\begin{eqnarray}}
\newcommand{\eeqs}{\end{eqnarray}}   
\numberwithin{equation}{section}
\newcommand{\be}{\begin{equation}}
\newcommand{\ee}{\end{equation}}
\newcommand{\bea}{\begin{eqnarray}}
\newcommand{\eea}{\end{eqnarray}}

\usepackage{graphicx}

\begin{document}

\title{From black strings to black holes:  \\
nuttier and squashed  AdS$_5$ solutions}

\author{{\large Yves Brihaye,}$^{\dagger}$
{\large Jutta Kunz }$^{\ddagger}$
and {\large Eugen Radu}$^{\ddagger}$ \\ 
$^{\dagger}${\small Facult\'e des Sciences, Universit\'e de Mons-Hainaut,
B-7000 Mons, Belgium}\\  
$^{\ddagger}${\small Institut f\"ur Physik, Universit\"at Oldenburg, Postfach 2503, D-26111 Oldenburg, Germany} }

\maketitle

%\ \ \ PACS Numbers: 04.50.+h, 11.10.Kk, 11.15.Kc

\bigskip

\begin{abstract}
We construct new solutions of the Einstein equations with negative
cosmological constant in five spacetime dimensions.
They smoothly emerge as deformations of the known AdS$_5$ black strings.
The first type of configurations can be viewed as the $d=4$ Taub-NUT-AdS 
solutions
uplifted to five  dimensions, in the presence of a negative
cosmological constant.
We argue that these solutions provide  the 
gravity dual for a ${\cal N}=4$ super-Yang-Mills theory formulated in a
$d=4$
homogeneous G\"odel-type spacetime background. 
A different deformation of the AdS$_5$ black strings leads to squashed AdS black holes and their topological
generalizations.
In this case, the conformal infinity is the product of time and a circle-fibration over
a base space that is a two-dimensional  Einstein space.

 \end{abstract}
\medskip
\medskip

 %%%%%%%%%%%%%%%%%%%%%%%%%%%%%%%%%%%%%%%%%%%%%%%%%%%%%%%%%%%%%%%%%%
\section{Introduction}
%%%%%%%%%%%%%%%%%%%%%%%%%%%%%%%%%%%%%%%%%%%%%%%%%%%%%%%%%%%%%%%%%%
Recently a tremendous amount of interest has been focused on 
 solutions of the Einstein equations in  more than $d=4$ spacetime
dimensions.
This interest was enhanced by the development of string theory, 
which requires a ten-dimensional spacetime,
to be consistent from a quantum point of view.

 Solutions with a number of compact dimensions, present for
$d\geq 5$ spacetime dimensions, are especially  interesting, since they 
exhibit new features that have no analogue in the usual $d=4$ theory.
For the case $d=5$ without a cosmological constant, 
the simplest configuration of this 
type is found by assuming translational symmetry along the 
extra coordinate direction and uplifting known solutions of the vacuum Einstein
equations in four dimensions. 
This corresponds to a vacuum uniform black string (UBS)
with horizon topology $S^{2}\times S^1$. It approaches asymptotically the 
four dimensional Minkowski-space times a circle,
the simplest case being the Schwarzschild black string.

Due to the absence of closed form solutions, relatively little is known
about the generalizations of these  configurations  
with a cosmological constant $\Lambda$.
The $d=5$ Anti-de Sitter (AdS) counterparts of the  Schwarzschild black string 
have been considered for the first time in 
 \cite{Copsey:2006br}.
Their generalizations to higher 
dimensions $d>5$ were discussed in \cite{Mann:2006yi},  
configurations with an event horizon topology 
$H^{d-3}\times S^1$ being considered as well. 
Various features of the AdS black strings\footnote{These solutions should not be confused with
the warped AdS configurations as discussed for instance in \cite{Chamblin:1999by}.
Although the warped solutions
are also sometimes called black strings in the literature, their properties are very different as compared to those
in \cite{Copsey:2006br},  \cite{Mann:2006yi}. }, including solutions with 
matter fields, can be found in 
\cite{Brihaye:2007vm}, \cite{Brihaye:2007jua}, \cite{Bernamonti:2007bu}.
As argued in \cite{Copsey:2006br, Mann:2006yi}, 
these solutions may have relevance in a 
AdS/CFT context, since they
provide the gravity dual of a field theory  on
a $S^{d-3}\times S^1\times R_t$ (or $H^{d-3}\times S^1\times R_t$)
background.

However, in four spacetime dimensions, the Schwarzschild black hole has
an interesting generalisation,
given by the famous Taub-NUT (TN) solution \cite{Taub:1951ez},\cite{NUT},\cite{Misner}.
This (Lorentzian signature-) metric has become renowned for being '{\it a counterexample to
almost anything}' \cite{misner-book}, and played an important role in conceptual
developments in general relativity \cite{Hawking}.
The TN solution  is usually interpreted as describing a gravitational
dyon with both ordinary and magnetic mass.
The nut charge $n$ plays there a dual role to ordinary  mass, 
in the same way that electric
and magnetic charges are dual within Maxwell theory \cite{dam}.
As discussed by many authors, 
the presence of magnetic-type mass
introduces a "Dirac-Misner-string singularity" in the metric (but no curvature singularity).
This can be removed by appropriate identifications and changes in 
the topology of the spacetime manifold, which imply a periodic time coordinate.  
The periodicity of the time coordinate prevents an interpretation 
of the TN metric as a usual black hole. 
Moreover, this metric is not asymptotically flat in the usual sense
although it does obey the required fall-off conditions (see \cite{Lynden-Bell:1996xj}
for an extended discussion of the properties of the TN solution).
The TN solution has also an interesting generalisation 
in heterotic string theory \cite{Johnson:2004zq}, with a full conformal field theory definition,
which possibly indicates 
that string
theory can very well live even in the presence of nonzero nut charge.

Of interest here are the generalisations of the four dimensional  TN solution 
with a cosmological constant,
which played an important role in (A)dS/CFT conceptual developments.
For example, the vacuum TNAdS solution in four dimensions provided the
first test bed for AdS/CFT correspondence in spacetimes where the asymptotic structure was only locally
asymptotic AdS \cite{Hawking:1998ct}, 
\cite{Chamblin:1998pz}, \cite{Emparan:1999pm}.

Similar to the Schwarzschild case, for $\Lambda=0$, the four dimensional vacuum TN solution can 
trivially be uplifted to extend into 
the extra spacelike direction, which leads to a  nuttier  black object.
However, this procedure cannot be repeated in the presence of a cosmological
constant and the issue of constructing  $d=5$ counterparts of the known four dimensional TNAdS solutions
has been scarcely explored in the literature. 
To our knowledge, the only known solutions possess the unusual
feature that there is a constraint between the possible values of the
nut charge and the cosmological constant \cite{Mann:2003zh}.

The main purpose of this work is to present a  new family of  solutions of $d=5$ vacuum
Einstein equations with a negative cosmological constant, which exist for 
any value of the cosmological constant.
These solutions emerge smoothly from the known AdS$_5$ black strings
 and represent natural generalisations of the $d=5$ AdS  "nuttier"  black holes in \cite{Mann:2003zh}.

Since we could not find a general closed form solution, the configurations in this paper are constructed numerically by matching the
 near-horizon expansion of the metric to their asymptotic Fefferman-Graham
form \cite{Feff-Graham}. Their global charges are computed by using the standard
 counterterm prescription.
 Our results indicate that the TNAdS$_5$ configurations share the basic properties
 of their $d=4$ counterparts.

Moreover, these solutions may teach us something
about the physics in spacetimes containing  closed timelike curves (CTCs).
Here we use the remark that
the boundary metric of these
TNAdS$_5$ solution is just the  two-parameter
family of $d=4$ G\"odel-type homogeneous spacetimes
 discussed in \cite{reboucas} 
 (the famous acausal G\"odel rotating universe \cite{Godel:1949ga}
representing a particular case).
Although the meaning of quantisation is 
unclear in the presence of CTCs, we argue that by
using the AdS/CFT correspondence, one may get an idea about the behaviour of a quantum field
theory in a such a background.

All configurations mentioned above have as basic building block
the $d=4$ TN solution with Lorentzian signature.
However, the main relevance of the $d=4$ nut-charged solutions is for a 
Euclidean signature of spacetime, in which case they have 
various applications in a quantum gravity context \cite{Hawking:ig}.
For $\Lambda=0$, the $d=5$ solution found 
by taking the product of a $d=4$ Euclideanized TN metric  with the time coordinate (a real line) 
represents the Gross-Perry-Sorkin  monopole \cite{Gross:1983hb,Sorkin:1983ns}, which
plays an important role in the context of Kaluza-Klein theories.
The Gross-Perry-Sorkin solution proves the existence there of
asymptotically {\it locally} flat configurations, approaching a twisted
$S^1$ bundle over a four dimensional Minkowski spacetime.
Black hole solutions with this type of asymptotics enjoyed recently some 
interest, following the discovery by
Ishihara and Matsuno (IM) \cite{Ishihara:2005dp} of a 
new charged solution in the five dimensional Einstein-Maxwell theory. 
The horizon  of the IM black hole has $S^3$ topology, while
its spacelike infinity is a squashed sphere or $S^1$ bundle over $S^2$.
This solution has been generalised in various directions, including configurations 
with more general gauge  fields \cite{Brihaye:2006ws}
and stability analysis \cite{Kimura:2007cr}.
  
The AdS counterparts of the IM squashed black holes were considered in the 
recent paper \cite{Murata:2009jt}. 
These solutions have a number of interesting properties, providing the gravity dual 
for a ${\cal N}=4$ super Yang-Mills theory on a  background whose spatial part is a squashed three sphere.

The second  purpose of this work is to present a detailed study of the AdS squashed black holes.
We argue that they can also be viewed as smoothly emerging from the AdS$_5$ black strings 
by turning on a squashing parameter $n$,
the usual Schwarzschild-AdS solution with a spherical horizon being found for a particular value of $n$.
New $d=5$ black hole solutions with a different topology of the event horizon are also presented.
In particular, we discuss the  asymptotic expansion of these configuration, compute their mass
and discuss their thermodynamical features.

The outline of this article is as follows. In the next Section we present the general framework while in Section 3 we
study the TNAdS$_5$ solutions. The AdS counterparts  of the IM squashed black holes 
are discussed in Section 4. We conclude in Section 5 with some final remarks.

%%%%%%%%%%%%%%%%%%%%%%%%%%%%%%%%%%%%%%%%%%%%%%%%%%%%%%%%%%%%%%%%%%
\section{The general framework}
%%%%%%%%%%%%%%%%%%%%%%%%%%%%%%%%%%%%%%%%%%%%%%%%%%%%%%%%%%%%%%%%%%
%%%%%%%%%%%%%%%%%%%%%%%%%%%%%%%%%%%%%%%%%%%%%%%%%%%%%%%%%%%%%%%%%%
\subsection{The action and the counterterm method}
%%%%%%%%%%%%%%%%%%%%%%%%%%%%%%%%%%%%%%%%%%%%%%%%%%%%%%%%%%%%%%%%%%
 We start with the following action in five spacetime 
dimensions
\begin{eqnarray}
\label{action}
I_0=\frac{1}{16 \pi G}\int_{\mathcal{M}} d^5 x \sqrt{-g}
 \left ( R-2 \Lambda \right)
-\frac{1}{8 \pi G}\int_{\partial\mathcal{M}} d^{4} 
x\sqrt{-\gamma}K,
\end{eqnarray}
where  $G$ is the gravitational constant,
$\Lambda=-6/ \ell^2$ is the cosmological constant,
 $\mathcal{M}$ is 
a five-dimensional manifold with metric $g_{\mu \nu }$, $K$ is the trace of the extrinsic 
curvature $K_{ij}=-\gamma_{i}^{k}\nabla_{k}n_{j}$ of 
the boundary $\partial M$ with unit normal $n^{j}$ and induced metric $\gamma_{ij}$.

As usual with AdS solutions,
it turns out that the action (\ref{action}) evaluated at the level of the equations of motion diverges. 
The general remedy for this situation is to add counterterms, 
\textit{i.e.} coordinate invariant functionals of the 
intrinsic boundary geometry that are specifically designed to cancel 
out the divergences\footnote{This technique is especially suited for the solutions
discussed in this paper which present no obvious background.}. 
%To regularize the divergences in the gravitational action sector, 
 Therefore, the following boundary counterterm part is added to the action (\ref{action})
\cite{Balasubramanian:1999re,Skenderis:2000in}:
\begin{eqnarray}
\label{ct}
I_{\mathrm{ct}}&=&-\frac{1}{8\pi G}\int_{\partial\mathcal{M}} d^{4}x\sqrt{-\gamma }
\bigg\{
-\left ( 
\frac{3}{\ell }+\frac{\ell }{4}{\cal R} 
\right )
+ \log(\frac{r}{\ell})
\left (  
 \frac{\ell^3}{8}
 (\frac{1}{3}{\cal R}^2-{\cal R}_{ij}{\cal R}^{ij}
)
\right )
\bigg\},
 \end{eqnarray}
where ${\cal R}$, $ {\cal R}_{ijkl} $ and ${\cal R}_{ij}$ are the curvature, Riemann 
and the Ricci tensor associated with the induced metric $\gamma_{ij}$.
The second term in   (\ref{ct}) is the usual expression required to cancel the
logarithmic divergence  that appears in $d=5$ dimensions
for some special boundary geometries,  $r$ being a coordinate normal to boundary.  
  
Using these counterterms, 
one can construct a divergence-free boundary stress 
tensor from the total action $I=I_0+I_{\mathrm{ct}}$ by defining a boundary stress-tensor: 
\begin{eqnarray}
\label{stress}
T^{ij}
=\frac 2{\sqrt{-\gamma}}\frac{\delta I  }{\delta\gamma_{ij}}
= 
\frac 1{8\pi G} \bigg \{K^{ij} - K\gamma^{ij} + \frac {\ell}{2}{\cal G}^{ij} 
 +  \log(\frac{r}{\ell})  \bigg [\frac{\ell^3}{8 } (\frac 13\gamma^{ij}{\cal R}^2 -
\gamma^{ij}{\cal R}_{kl}{\cal R}^{kl} 
\\
\nonumber
- \frac{4}{3}{\cal R}{\cal R}^{ij} 
+ 4{\cal R}^{ikjl}{\cal R}_{kl} 
+2\square\lp{\mathcal R}^{ij}-\frac12\ga^{ij}{\mathcal R}\rp
+\frac23\lp\ga^{ij}\square-\nabla^i\nabla^j\rp{\mathcal R}
 ) 
 \bigg]
 \bigg\},
 \end{eqnarray}
 (with ${\cal G}^{ij}$ the Einstein tensor of the boundary metric).
% Following the standard prescription, we write the metric on the boundary in the form:
%\beqs
%\gamma_{ij}dx^idx^j=-N^2dt^2+\sigma_{ab}(dy^a+N^adt)(dy^b+N^bdt),
%\label{ADMboundary}
%\eeqs
%where $N$ and $N^a$ are the lapse function, respectively 
%the shift vector, and $y^a$, $a=1,2,3$ are the intrinsic 
%coordinates on a closed surface $\Sigma$ of constant time 
%$t$ on the boundary.  
Then a conserved charge 
associated with a Killing vector $\xi ^{i}$ at infinity can be calculated using the
relationship 
\begin{equation}
{\frak Q}_{\xi }=\oint_{\Sigma }d^{3}S^i \xi ^{j}T_{ij},
\label{Mcons}
\end{equation}%
where $\Sigma $ is a closed surface. 
%(with normal $u^{i}$). 
The conserved mass/energy $M$ is the charge associated 
with the time translation symmetry, with $\xi =\partial /\partial t$.

Typically the boundary of this kind of spacetime will be an asymptotic
surface at some large radius $r$. However the metric restricted to the
boundary $\gamma _{ij}$ diverges due to an infinite conformal factor $%
r^{2}/\ell ^{2}$, and so the metric upon which the dual field theory resides
is defined using the rescaling 
\begin{eqnarray}  \label{resc}
h_{ij}=\lim_{r\rightarrow \infty }\frac{\ell ^{2}}{r^{2}}\gamma _{ij}.
\end{eqnarray}%
The stress-energy tensor $%
<\tau _{ij}>$ for the dual theory formulated in a background metric given by $h_{ij}$,
  can be calculated by using the following
relation \cite{Myers:1999qn} 
\begin{eqnarray}  
\label{r1}
\sqrt{-h}h^{ik}<\tau _{kj}>=\lim_{r\rightarrow \infty }\sqrt{-\gamma }\gamma
^{ik}T_{kj}.
\end{eqnarray}

%%%%%%%%%%%%%%%%%%%%%%%%%%%%%%%%%%%%%%%%%%%%%%%%%%%%%%%%%%%%%%%%%%
\subsection{Uniform black strings in AdS$_5$}
%%%%%%%%%%%%%%%%%%%%%%%%%%%%%%%%%%%%%%%%%%%%%%%%%%%%%%%%%%%%%%%%%%
Since all solutions discussed in this paper are connected with the UBS in 
\cite{Copsey:2006br, Mann:2006yi},
we briefly present here their basic properties.
A convenient metric ansatz to study this type of configurations is
\begin{eqnarray}
\label{metric-UBS} 
ds^2= \frac{dr^2}{f(r)}+r^2 d\Omega_k^2
%(d\theta^2+F_k^2(\theta) d\varphi^2)
+a(r)dz^2-b(r)dt^2, 
\end{eqnarray}
where $d\Omega_k^2=d\theta^2+F_k^2(\theta) d\varphi^2$
is the metric on a two-dimensional surface of constant curvature $2k$.
The discrete parameter $k$ takes the values $1,0,-1$ and implies the form of
the function $F_{k}(\theta )$:
\begin{eqnarray}  
\label{Fk}
F_{k}(\theta )=\left\{ 
\begin{array}{ll}
\sin \theta , & \mathrm{for}\ \ k=1 \\ 
\theta , & \mathrm{for}\ \ k=0 \\ 
\sinh \theta , & \mathrm{for}\ \ k=-1.%
\end{array}%
\right.
\end{eqnarray}%
Thus, for $k=1$, $\theta$ and $\varphi$ are the spherical coordinates with the usual range;
for $k=0,-1$ although $\varphi$ is a periodic coordinate with $0\leq \varphi \leq 2 \pi$,
the range of $\theta$ is not restricted.  
When $k = 0$, a constant $(r,t)$ slice is a flat surface,
while for $k = -1$, this sector is a space with constant negative curvature, also known as a hyperbolic
plane.
In what follows, $V_{k}$ will denote the total area of the ($\theta,\varphi)$ surface.
As usual with black strings, the coordinate along the compact direction is denoted by $z$ and its
asymptotic length is $L$ (the value of $L$ is arbitrary for the UBSs).
%$r$ is the radial coordinate, with $r\geq r_h$.
The event horizon of a black string is located at $r=r_h$,
where $f(r_h)=b(r_h)=0$ and $a(r_h)>0$.
On a basic conceptual level, these solutions can be viewed as the four dimensional
Schwarzschild-AdS$_4$ solutions `uplifted' to $d=5$ in the presence of
a negative $\Lambda$.
Different from the $\Lambda=0$ case,  the limit $r_h\to 0$ of the AdS black 
string solutions  with an event horizon 
topology $S^{d-3}\times S^1$ 
corresponds to a nontrivial globally regular, soliton-like configuration. 
%have a nontrivial, 
%globally regular 
%limit with zero event 
%horizon radius.

The equations satisfied by the metric functions $a,b,f$ are presented $e.g.$
in \cite{Mann:2006yi}.
Unfortunately no exact solution is known in the vacuum case\footnote{Exact 
solutions have been found in  Einstein-Maxwell-$\Lambda$ theory \cite{Bernamonti:2007bu},
\cite{Chamseddine:1999xk} for special values of the gauge coupling constant.} and thus
one has to resort to numerical techniques.
The expression of the solutions in the near horizon region and for large $r$ can be 
found by taking $n=0$
in the relations (\ref{eh-nut}), (\ref{as-NUT}) below.

Finally, by performing the double analytic continuation $t\to i \chi$, $z \to i \tau$,
the black strings become static bubbles of nothing, with a line element
\begin{eqnarray}
\label{bubble} 
ds^2= \frac{dr^2}{f(r)}+r^2(d\theta^2+F_k^2(\theta) d\varphi^2)+b(r)d\chi^2-a(r)d\tau^2. 
\end{eqnarray}
The $S^1$ factor of the metric pinches off at the radius $r_h$, and the solution is regular if the spatial 
coordinate $\chi$ is identified with a period $\beta=1/T_H$ (with $T_H$ the
Hawking temperature of the initial black string solution).

%%%%%%%%%%%%%%%%%%%%%%%%%%%%%%%%%%%%%%%%%%%%%%%%%%%%%%%%%%%%%%%%%%
\subsection{The Taub-NUT-AdS$_4$ solution}
%%%%%%%%%%%%%%%%%%%%%%%%%%%%%%%%%%%%%%%%%%%%%%%%%%%%%%%%%%%%%%%%%%
For completeness, we briefly discuss here also the basic properties of the nut-charged Lorentzian AdS$_4$ solutions.
%For $\Lambda =-3/\ell^2<0$, it has been shown 
%\cite{Chamblin:1998pz} that the TN spacetime can be generalized by including an
%additional discrete parameter $k$ . The case $k\neq 1$ corresponds to
%solutions where the angular spheres $(\theta ,\varphi $) are replaced by
%planes $(k=0)$ or hyperboloids ($k=-1$).
For $\Lambda=0$, the usual Taub-NUT construction corresponds to a $U(1)$-fibration 
over a two-dimensional Einstein space used as the base space.
Usually taken to be a sphere, for $\Lambda =-3/\ell^2<0$  this space can also be a torus or
an hyperboloid \cite{Chamblin:1998pz}.
For a metric ansatz similar to that used in what follows for $d=5$, 
the TNAdS$_4$ is given by\footnote{The usual form of the metric used in the literature is recovered by taking
$r\to\sqrt{r^2+n^2}$.} 
\begin{eqnarray}
\label{metric-NUT4} 
ds^2= \frac{dr^2}{f(r)}+r^2(d\theta^2+F_k^2(\theta) d\varphi^2)-b(r)(dt+4 n F_k^2(\frac{\theta}{2})d\varphi)^2,
\end{eqnarray}
where
\begin{eqnarray}
\label{mf-NUT4} 
&b(r)=k\left(1-\frac{2n^2}{r^2} \right)+\frac{1}{r^2}\left(-2M \sqrt{r^2-n^2}+\frac{1}{\ell^2}(r^4+4n^2r^2-8n^4)\right),~ 
f(r)=\left(1-\frac{n^2}{r^2}\right)b(r),
\end{eqnarray}
$M$ being a parameter fixing the mass of solutions.
For $k=1$, in order to avoid a conical singularity at $\theta=\pi$, the coordinate $t$
must be identified with period $8\pi n$ which yields a spacetime with CTCs.
Although less obvious, all $k=0$ solutions and the $k=-1$ metrics with 
$4n^2> \ell^2$ are also not globally hyperbolic \cite{Kerner:2006vu}, \cite{Astefanesei:2004kn}.

It is also instructive to give the large $r$ asymptotics of the metric functions  in (\ref{metric-NUT4})
\begin{eqnarray}
\label{as-NUT4} 
b(r)=\frac{r^2}{\ell^2}+(k+\frac{4n^2}{\ell^2})-\frac{2M}{r}+O(1/r^2),~~
f(r)=\frac{r^2}{\ell^2}+(k+\frac{3n^2}{\ell^2})-\frac{2M}{r}+O(1/r^2).
\end{eqnarray}
 In the limit of vanishing nut
charge, the metric (\ref{metric-NUT4}) describes topological black holes.

%%%%%%%%%%%%%%%%%%%%%%%%%%%%%%%%%%%%%%%%%%%%%%%%%%%%%%%%%%%%%%%%%%
\section{Generalized Taub-NUT-AdS$_5$ solutions}
%%%%%%%%%%%%%%%%%%%%%%%%%%%%%%%%%%%%%%%%%%%%%%%%%%%%%%%%%%%%%%%%%%
%%%%%%%%%%%%%%%%%%%%%%%%%%%%%%%%%%%%%%%%%%%%%%%%%%%%%%%%%%%%%%%%%%
\subsection{The ansatz and asymptotics}
%%%%%%%%%%%%%%%%%%%%%%%%%%%%%%%%%%%%%%%%%%%%%%%%%%%%%%%%%%%%%%%%%% 
Unfortunately, there is no prescription to uplift a four dimensional solution to higher dimensions
in the presence of a cosmological constant.
However, the expressions (\ref{metric-UBS}), (\ref{metric-NUT4}) above, 
naturally lead to the following metric ansatz for 
TNAdS$_5$ solutions: 
\begin{eqnarray}
\label{metric-NUT} 
ds^2= \frac{dr^2}{f(r)}+r^2(d\theta^2+F_k^2(\theta) d\varphi^2)+a(r)dz^2
-b(r)(dt+4 n F_k^2(\frac{\theta}{2})d\varphi)^2,
\end{eqnarray}
with $k=\pm 1,0$ and $F_k(\theta)$ still given by (\ref{Fk}).
The range of the coordinates $\theta,\varphi$ and $z$ for this line element is similar to the black string case case.
 
The Einstein equations with a negative cosmological constant imply 
that the metric functions $a(r)$, $b(r)$ and $f(r)$ are solutions of 
the following equations:
\begin{eqnarray}
\nonumber
&&f'=\frac{2k}{r}
+\frac{8r}{\ell^2}
-\frac{2f}{r}
-f\left(\frac{a'}{a}+\frac{b'}{b} \right)+\frac{4n^2 b}{r^3}~,
\\ 
\label{eqs-NUT} 
&&a''=\frac{2a}{r^2} 
-\frac{2k a}{r^2f}
-\frac{4a}{ \ell^2f}
+\frac{2ab'}{rb}
+\frac{ a'}{r}
-\frac{k a'}{rf}
-\frac{4 ra'}{\ell^2f}
+\frac{a'b'}{2b}
+\frac{a'^2}{a}-\frac{2n^2b (a+r a')}{r^4 f}~,
\\
\nonumber
&&\frac{b'}{b}= 2\frac{ a\big[ 2\ell^2(k-f)+12r^2\big]
-2 r\ell^2fa'}{r\ell^2f\big[ra'+4 a\big]}+\frac{4n^2ab}{ r^3 f(r a'+4a)}.
\end{eqnarray}
%For $n=0$, the ansatz (\ref{metric-NUT}) and the field equations (\ref{eqs-NUT})
%reduce to those employed in  \cite{ Mann:2006yi} for the AdS uniform black string case. 
We are interested in solutions with a nonextremal horizon located at $r=r_h$.
As $r \to r_h$, the following approximate form of the metric functions holds:
\begin{eqnarray}
\nonumber
&&a(r)=a_h+a_1 (r-r_h)+a_2 (r-r_h)^2+O(r-r_h)^3,~~
b(r)= b_1 (r-r_h)+b_2 (r-r_h)^2+O(r-r_h)^3,
\\
\label{eh-nut}
&&f(r)=f_1 (r-r_h)+f_2 (r-r_h)^2+O(r-r_h)^3,~~
\end{eqnarray}
in terms of two positive parameters $b_1$, $a_h$. One finds $e.g.$
\begin{eqnarray}
\label{exp3} 
 f_1=\frac{k}{r_h }+\frac{4r_h}{\ell^2 },~~a_1=\frac{8a_hr_h}{4r_h^2+k\ell^2},~~
 f_2=\frac{3b_1 n^2} {2 r_h^3}-\frac{k}{r_h^2}-\frac{2}{\ell^2},
 \\
 \nonumber
 a_2=\frac{4a_h(4 r_h^3-b_1 n^2\ell^2)}{r_h(4r_h^2+k \ell^2)^2},~~
 b_2=-\frac{b_1(4r_h^3+b_1n^2\ell^2+2k r_h\ell^2)}{8r_h^4+2k r_h^2\ell^2}
 .
\end{eqnarray}
The condition for a regular 
  horizon is $f'(r_h)>0$, which for $k=-1$
implies the existence of a minimal value of $r_h$, $i.e.$
$
r_h>\ell/2.
$

At large $r$, the functions appearing in the metric admit the following
 expansion:
\begin{eqnarray}
\nonumber
&&a(r)=\frac{r^2}{\ell^2}+\frac{1}{2}(k+\frac{2n^2}{\ell^2}) 
+\frac{\ell^2}{r^2}\left(c_z+ \frac{1}{12}(k+\frac{4n^2}{\ell^2})(k+\frac{8n^2}{\ell^2}) 
\log(\frac{r}{\ell})\right)+O\left(\frac{\log r}{r^4}\right),
\\
\label{as-NUT} 
&&b(r)=\frac{r^2}{\ell^2}+\frac{1}{2}(k+\frac{4n^2}{\ell^2}) 
+\frac{\ell^2}{r^2}\left(c_t+ \frac{1}{12}(k+\frac{4n^2}{\ell^2})(k+\frac{16n^2}{\ell^2}) 
\log(\frac{r}{\ell})\right)+O\left(\frac{\log r}{r^4}\right),
\\
\nonumber
&&f(r)=\frac{r^2}{\ell^2}+\frac{2}{3} (k +\frac{5n^2}{2 \ell^2})  
+\frac{\ell^2}{r^2}\left(c_t+c_z-\frac{n^2}{2\ell^2}(k+\frac{4n^2}{\ell^2}) 
+\frac{1}{6}(k+\frac{4n^2}{\ell^2})(k+\frac{12n^2}{\ell^2}) 
\log(\frac{r}{\ell})\right)+O\left(\frac{\log r}{r^4}\right),
\end{eqnarray}
which generalizes for $n\neq 0$ the expansion derived in \cite{Mann:2006yi},  \cite{Bernamonti:2007bu}
(note the occurence here of
$\log$ terms even for the planar case $k=0$).

Remarkably, the equations (\ref{eqs-NUT} ) admit an exact solution found by Mann and Stelea 
in \cite{Mann:2003zh} for $n^2=\ell^2/4$, $k=-1$
(this being the only case when the log terms are absent in the asymptotics (\ref{as1})):
\begin{eqnarray}
\label{Mann-Stelea} 
f(r)=\frac{r^2}{\ell^2}-\frac{1}{4}+\frac{\ell^2}{r^2}c_t-\frac{\ell^4}{4 r^4}c_t,~~~
a(r)=\frac{r^2}{\ell^2}-\frac{1}{4},~~~
b(r)=\frac{r^2}{\ell^2} +\frac{\ell^2}{r^2}c_t.
\end{eqnarray}
The horizon of this solution is at $r_h=\ell c_1$ (where $c_t=-c_1^4)$ and has $a_h=c_1^2-1/4$ (this implies
$c_1>1/2$).

%%%%%%%%%%%%%%%%%%%%%%%%%%%%%%%%%%%%%%%%%%%%%%%%%%%%%%%%%%%%%%%%%%
\subsection{Global charges and general features}
%%%%%%%%%%%%%%%%%%%%%%%%%%%%%%%%%%%%%%%%%%%%%%%%%%%%%%%%%%%%%%%%%%
For $n$ different from zero, the $k=1$ metric structure (\ref{metric-NUT}) 
shares the same troubles 
exhibited by usual $d=4$ TN metric, $i.e.$
the solutions cannot be interpreted properly as black holes.
Requiring the absence of singularities due to the  Misner string implies  
a periodicity $8\pi n$ for the time coordinate.
For $k=0,-1$ the fibration is trivial and there is no Misner string singularity and no required periodicity for $t$.
However, the pathology of CTCs still occurs for all $k=0$ metrics
and some of the $k=-1$ solutions.
This can be seen $e.g.$ by looking at the $g_{\varphi \varphi}$
component of the metric tensor
\begin{eqnarray}  
\label{gfifi}
g_{\varphi \varphi}=r^2 F_k^2(\theta)-16 b(r) n^2 F_k^4(\theta/2)=
4  F_k^2(\theta/2)\left(r^2(1-k F_k^2(\theta/2))-4 n^2 b(r)F_k^2(\theta/2) \right).
\end{eqnarray}%
For $k=0$, $g_{\varphi \varphi}$ becomes negative for large enough 
values of $\theta$.
Thus the integral curves of the Killing vector $\partial/\partial \varphi$ provide an example of closed curves for 
a timelike motion.
The case $k=-1$ is more involved, the sign of $g_{\varphi \varphi}$ being fixed by the
ratio $4n^2 b(r)/r^2$.
Although we could not prove that rigorously, 
our numerical results suggest that the  value $4n^2/\ell^2$
($i.e.$ the Mann-Stelea
solution (\ref{Mann-Stelea}))
 always
separates the $k=-1$ geometries with CTCs from those which admit a global time coordinate.
(We recall that this is also the critical value for the TNAdS$_4$ metric 
(\ref{metric-NUT4}), separating the 
$k=-1$ globally hyperbolic solutions from those presenting CTCs).

Although there is no clear definition of the  global charges, Hawking temperature
and  entropy in the presence of CTCs, 
one can formally extend some results valid in the $k=-1$ globally hyperbolic case
to the general situation (see $e.g.$ the related discussion in \cite{Kerner:2006vu,Mann:2004mi}).

Thus, as for UBSs, the nuttier solutions would possess two global charges--the 
mass $M$ and tension ${\cal T}$, associated with the Killing vectors $\partial/\partial t$
and $\partial/\partial z$, respectively.
Their values are fixed by the constants $c_t$ and $c_z$, which enter  
 the expression of the metric functions at infinity (\ref{as-NUT}). 
To compute $M$ and  ${\cal T}$ we use the prescription (\ref{Mcons}), with the result 
\begin{eqnarray}
\nonumber
M&=&M_0+M_c^{(k)}~,
{\rm where}~M_0=\frac{\ell }{16\pi G 
}\big[c_z-3c_t\big]LV_{k},
~{\rm and}~
M_c^{(k)}=
\frac{V_{k}L\ell}{192\pi G}\left(k^2+\frac{19n^2}{\ell^2}(\frac{7n^2}{\ell^2}+2k) \right),
\\
\label{MT-nut} 
{\mathcal T}&=&{\mathcal T}_0+{\mathcal T}_c^{(k)}~,
~~{\rm where}~~~{\mathcal T}_0=\frac{\ell }{16\pi G 
}\big[3c_z-c_t\big] V_{k},
~~{\rm and}~~
{\mathcal T}_c^{(k)}=-
\frac{V_{k}\ell}{192\pi G} (\frac{n^2}{\ell^2}+k)^2.
\end{eqnarray}  
$M_c^{(k)}$ and ${\mathcal T}_c^{(k)}$ in the above relations are Casimir-like terms which appear also
in the $k=\pm 1$ black string limit.

The Hawking temperature of these solutions, as computed from the surface gravity is 
$T_H=\sqrt{f'(r_h)b'(r_h)}/4\pi$.
Of interest is also the horizon area, given by $A_H=r_h^2 \sqrt{a(r_h)} V_k L$. 
The entropy of the solutions can be identified with one quarter of $A_H$
for globally hyperbolic solutions only ($i.e.$ $k=-1,n^2\leq \ell^2/4$).
To our knowledge, the  
 thermodynamics of a (Lorentzian signature--) acausal spacetime is an open problem.

%%%%%%%%%%%%%%%%%%%%%%%%%%%%%%%%%%%%%%%
\begin{figure}[ht]
\hbox to\linewidth{\hss%
	\resizebox{7cm}{6.1cm}{\includegraphics{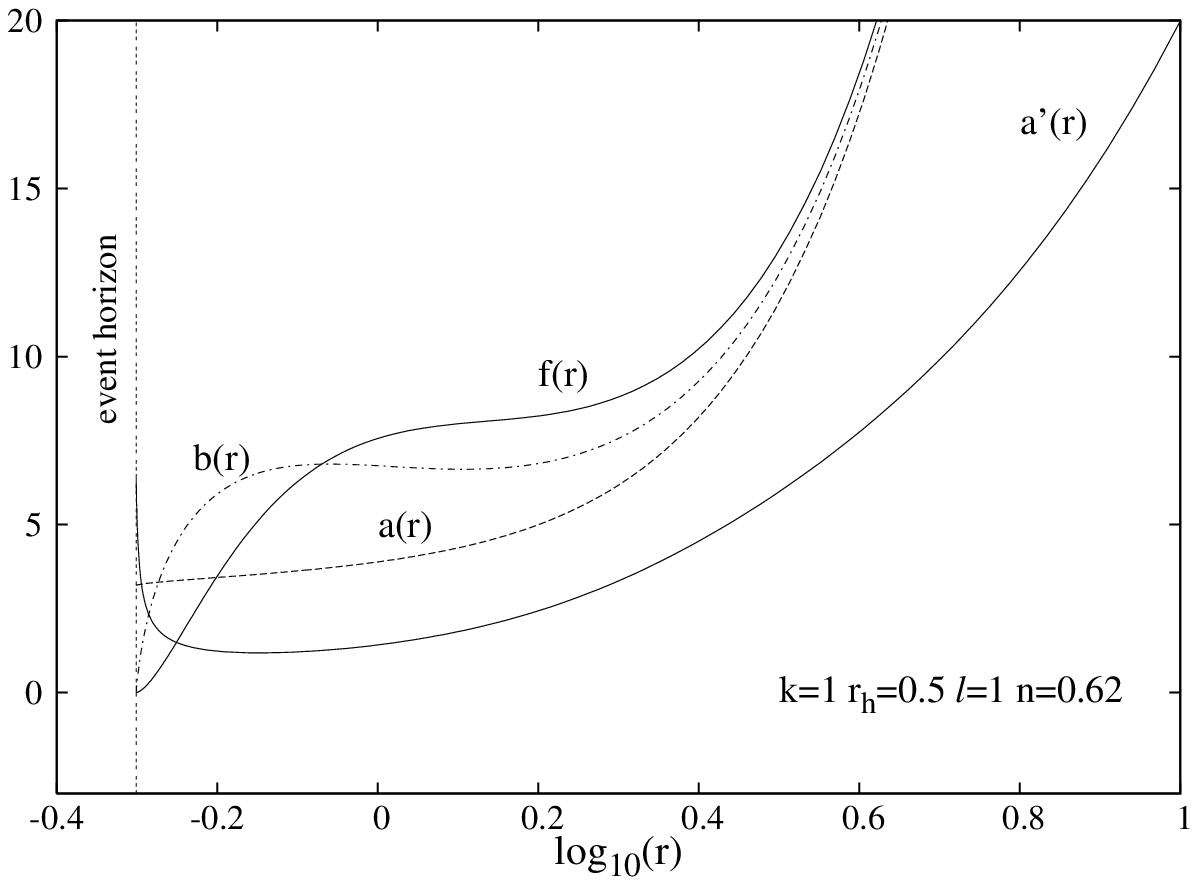}}
\hspace{5mm}%
        \resizebox{7cm}{6.1cm}{\includegraphics{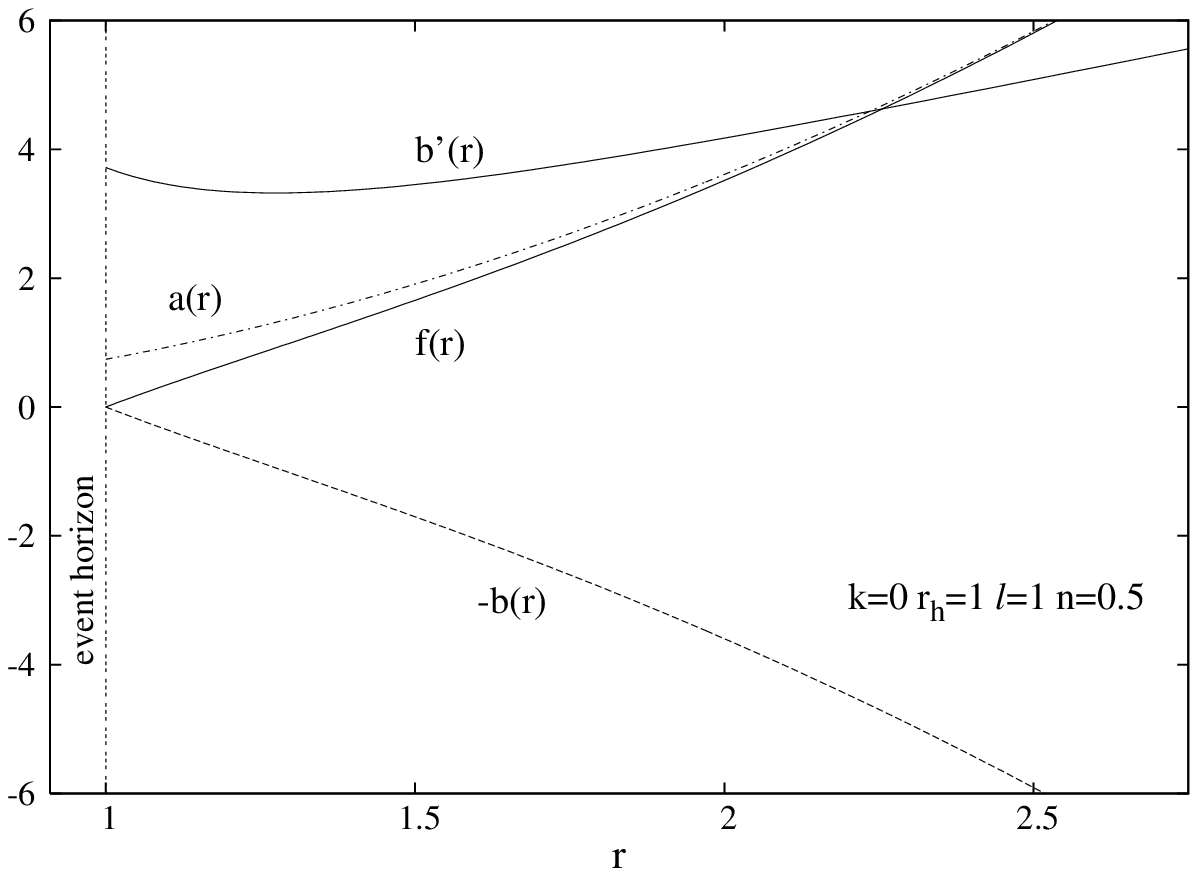}}	
\hss}
	\caption{  The profiles of typical $k=1,0$ nuttier solutions. 
} 
\label{fig_neg_n2_a}
\end{figure}
%%%%%%%%%%%%%%%%%%%%%%%%%%%%%%%%%%%%%%% 
%%%%%%%%%%%%%%%%%%%%%%%%%%%%%%%%%%%%%%%%%%%%%%%%%%%%%%%%%%%%%%%%%%
\subsection{Numerical results}
%%%%%%%%%%%%%%%%%%%%%%%%%%%%%%%%%%%%%%%%%%%%%%%%%%%%%%%%%%%%%%%%%% 
%
In the absence of explicit solutions for the generic values of the parameters,
we have solved the system of equations (\ref{eqs-NUT}) numerically. 
The computation of mass and tension for these solutions 
is a nontrivial problem which requires a very good numerical accuracy,
since the coefficients $c_t,c_z$
appear as subleading terms in the asymptotic
expansion (\ref{as-NUT}) (see the Ref. \cite{Bernamonti:2007bu} for a detailed discussion of 
this point). 

Following the methods in \cite{Mann:2006yi},\cite{Brihaye:2007vm}, 
we have looked for solutions which have a regular horizon at $r=r_h$, 
treating the equations as a boundary value problem for $r \in [r_h,\infty]$ 
(thus we did not consider
the behaviour of the solutions inside the horizon).
The Einstein equations were solved
by employing a collocation
 method for 
boundary-value ordinary
differential equations, equipped with an adaptive mesh selection procedure
\cite{colsys}.
Typical mesh sizes include $10^3-10^4$ points.
The solutions have a typical relative accuracy of $10^{-8}$.  

 Our numerical results clearly indicate the existence for any $k$
of solutions of the equations (\ref{eqs-NUT}), smoothly interpolating
 between the asymptotics (\ref{eh-nut}) and (\ref{as-NUT}).
 We have also verified that the solutions are singularity free for $r\geq r_h$.
In particular, the Kretschmann scalar stays finite everywhere.
 In this approach, the input parameters are $n$, $\ell$ and $r_h$.
 The event horizon data $a(r_h)$, $b'(r_h)$ and the coefficients
 at infinity $c_t$ and $c_z$ are read from
 the numerical output.

 %%%%%%%%%%%%%%%%%%%%%%%%%%%%%%%%%%%%
\begin{figure}[ht]
\hbox to\linewidth{\hss%
	\resizebox{7cm}{6.1cm}{\includegraphics{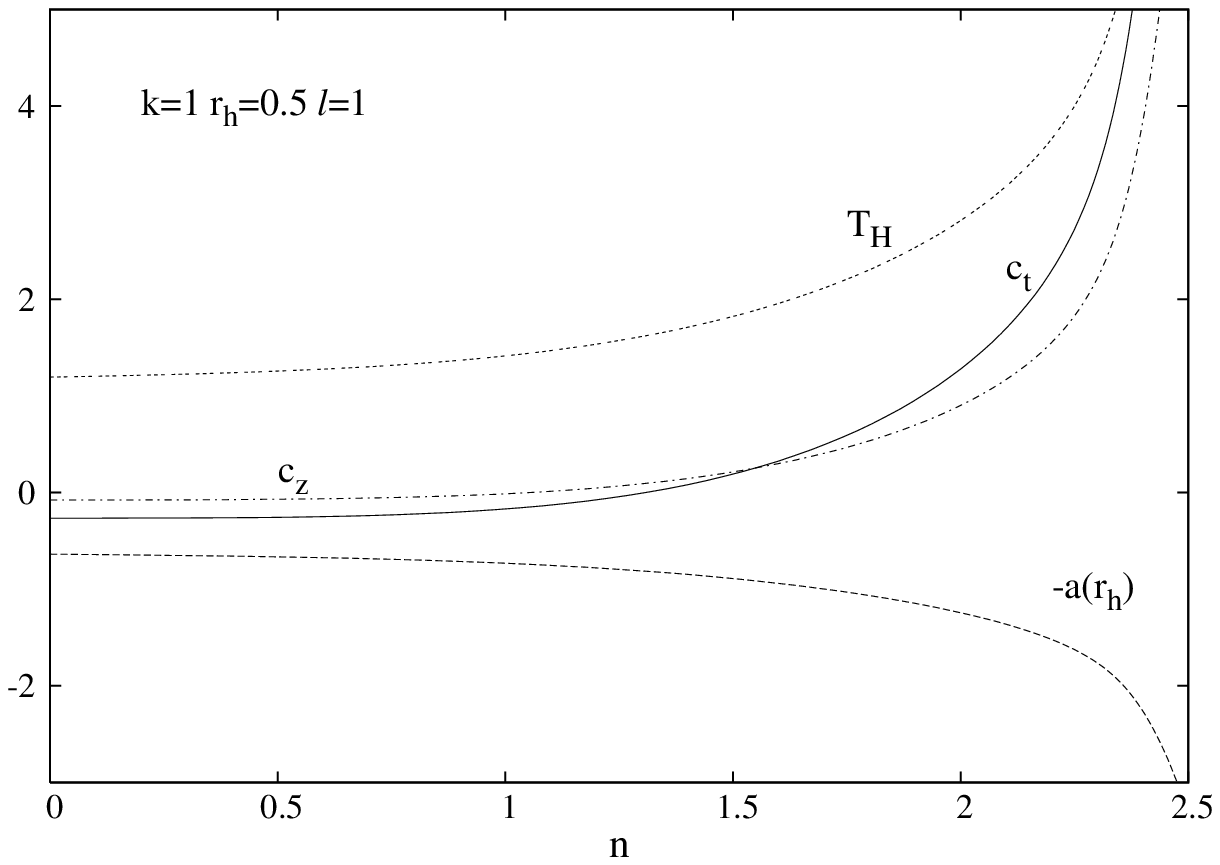}}
\hspace{5mm}%
        \resizebox{7cm}{6.1cm}{\includegraphics{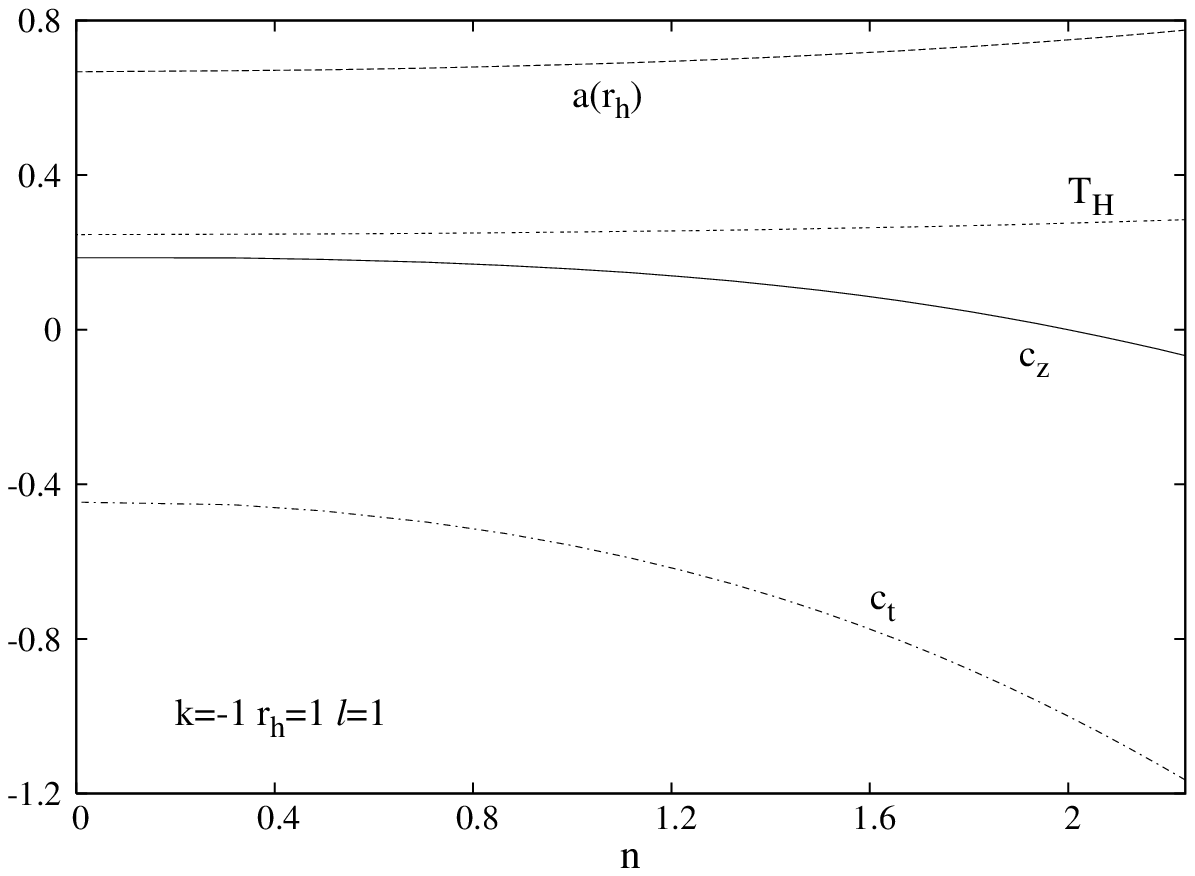}}	
\hss}
	\caption{  The dependence of the $k=1,-1$ nuttier solutions on the parameter $n$ for fixed 
	event horizon radius $r_h$.
} 
\label{Fig1}
\end{figure}
%%%%%%%%%%%%%%%%%%%%%%%%%%%%%%%%%%%%%%%

%%%%%%%%%%%%%%%%%%%%%%%%%%%%%%%%%%%%
\begin{figure}[ht]
\hbox to\linewidth{\hss%
	\resizebox{7cm}{6.1cm}{\includegraphics{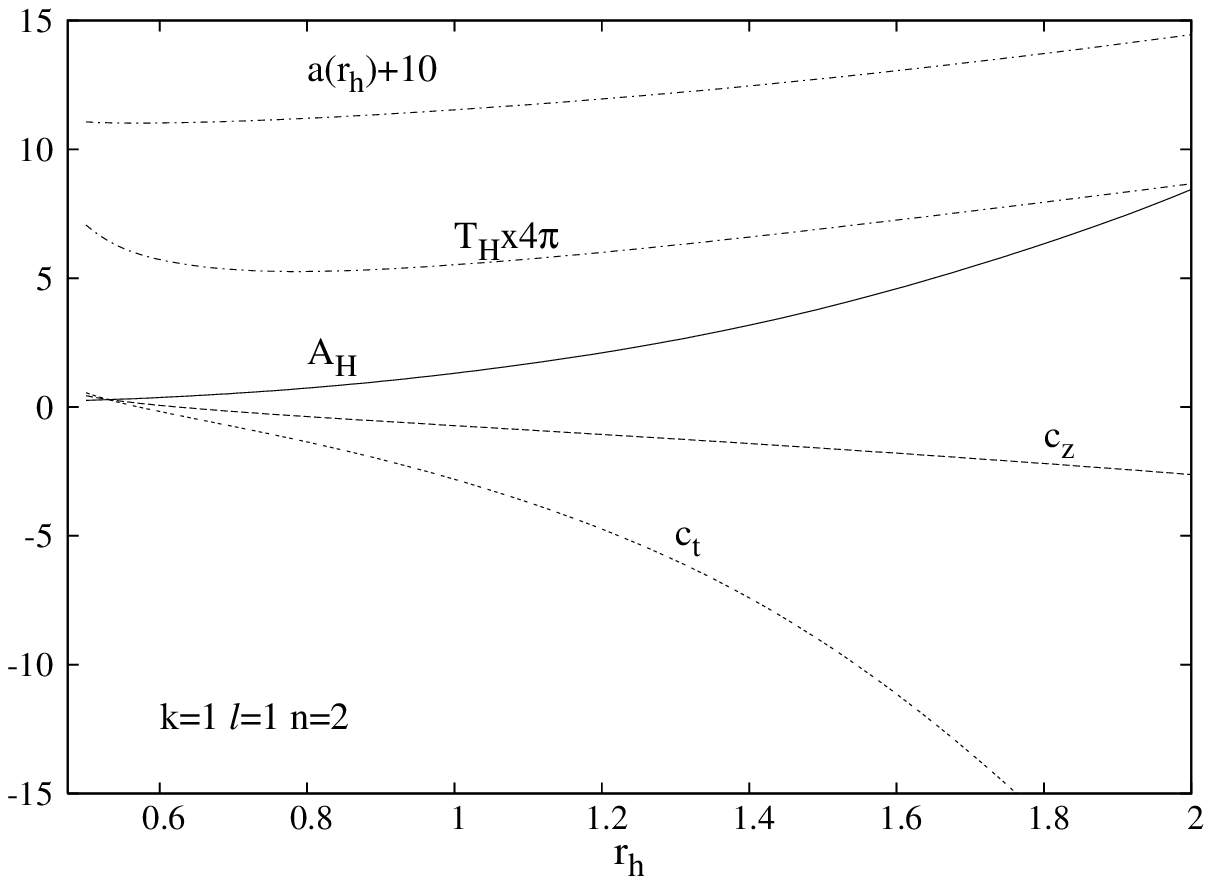}}
\hspace{5mm}%
        \resizebox{7cm}{6.1cm}{\includegraphics{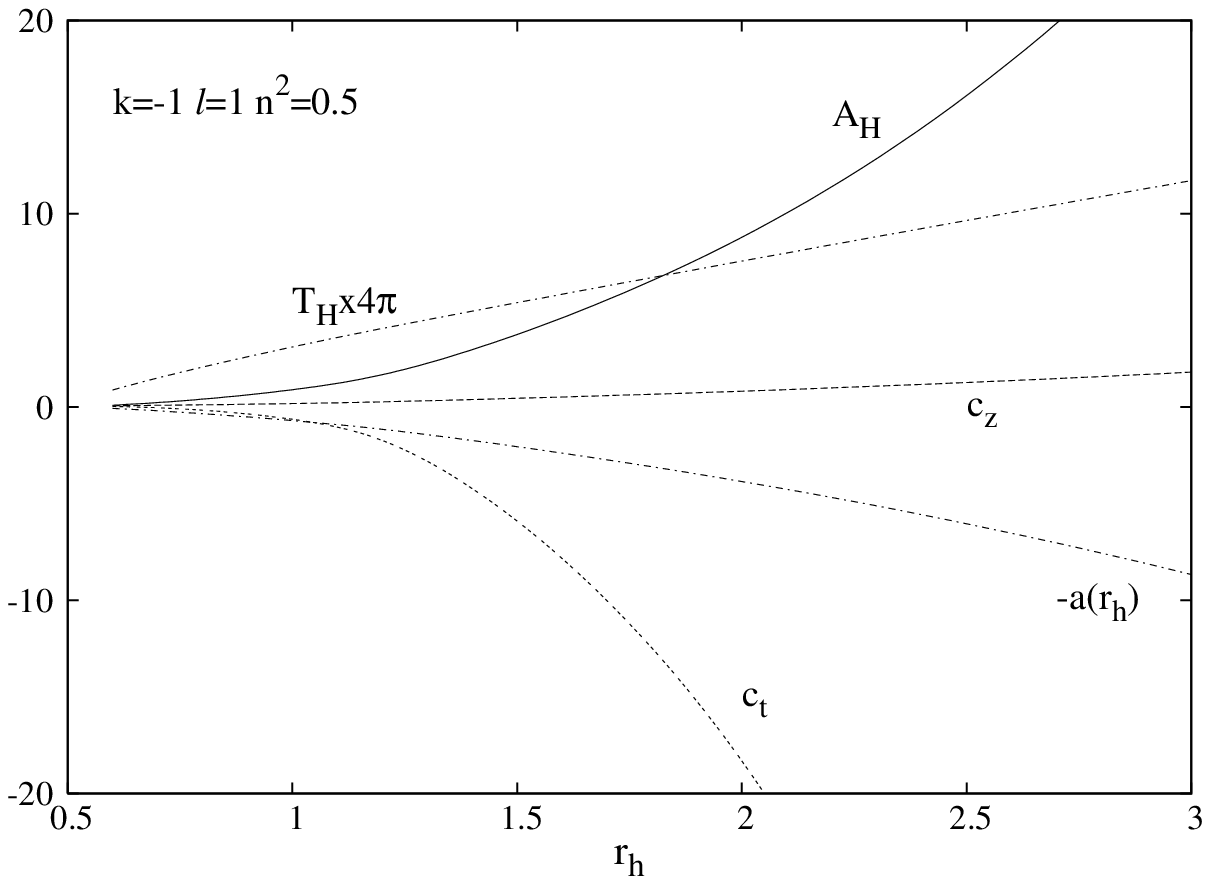}}	
\hss}
	\caption{  The dependence of the solutions on the event horizon radius $r_h$ for $k=\pm 1$ nuttier solutions with fixed 
	 $n$ .
} 
\label{Fig1}
\end{figure}
%%%%%%%%%%%%%%%%%%%%%%%%%%%%%%%%%%%%%%%

In practice we fix  the AdS length scale $\ell = 1$ and construct families of solutions by varying 
$n$ and $r_h$ for the three possible values of $k$. 
The choice $\ell=1$ does not affect the generality of the results,
since the
cosmological constant can be arbitrarily rescaled by an appropriate redefinition of the radial variable 
$r$ and of the parameter $n$.
Also, since only $n^2$ appear in the equations, we shall restrict to positive values of $n$.

The solutions with $n=0$ correspond to the UBSs discussed in \cite{Copsey:2006br},
\cite{Mann:2006yi} and were used as initial guess in the numerical iteration. 
As typical examples of 
nut-charged solutions, we plot in Figure $1$ 
two solutions with $k=1,0$. 

When increasing $n$, our numerical results show that
the uniform black string gets continuously deformed. 
For a given value of the event horizon radius, the coefficients
at infinity $c_t,c_z$ as well as the horizon 
parameters $T_H$ and $a(r_h)$  increase 
monotonically with $n$.
The corresponding picture in this case is shown in Figure 2
for both $k=1$ and $k=-1$ solutions (the picture for $k=0$ does not 
present new qualitative features).
For the $k=1$ case, when we continue to increase  $n$, the numerical results clearly
show that the derivative  of the function $b$ at the horizon diverges 
(while the derivative of the function $f$
remains constant) when a maximal value  $n = n_c$ is approached. 
As a consequence,  the surface gravity diverges in the limit $n \to n_c$.
A similar maximal value of $n$ is likely to exist also for $k=0,-1$.
However, in this case it has proven difficult to compute accurately the asymptotic
coefficients $c_t$, $c_z$ for large $n$.

As expected, for any value of $n\neq 0$, we notice also the presence of 
a minimal value $r_h^{(m)}$ of the event horizon radius.
As $r_h \to r_h^{(m)}$ a critical solution is approached
and the numerical solver fails to converge.
The study of this critical 
solution seems to require a different
parametrization of the metric ansatz and is beyond the purposes of this paper. 
 Also, for any $k$ and $n$, we did not find 
 a maximal allowed value of the event horizon radius.
  
 %%%%%%%%%%%%%%%%%%%%%%%%%%%%%%%%%%%%%%%%%%%%%%%%%%%%%%%%%%%%%%%%%%
\subsection{The boundary metric and the dual CFT}
%%%%%%%%%%%%%%%%%%%%%%%%%%%%%%%%%%%%%%%%%%%%%%%%%%%%%%%%%%%%%%%%%% 
The line element of
the boundary metric for the $d=5$ TN solutions  (\ref{metric-NUT})
is found by employing the prescription (\ref{resc}) and reads
\begin{eqnarray}  
\label{bound-gen}
ds^{2}=\ell ^{2}\left(d\theta ^{2}+F_{k}^{2}(\theta)d\varphi^{2}\right)
+dz^2
-\Big(4 n F_{k}^{2}(\frac{\theta}{2})d\varphi +dt\Big)^{2}.
\end{eqnarray}%  
 The relation (\ref{r1}) predicts the following form of the stress tensor 
 of the ${\cal N}=4$ super-Yang-Mills theory formulated in 
 this background, consisting in two parts:
\begin{eqnarray}
\label{tik-nut}
<\tau_i^{j}>=  <\tau_i^{j(f)}> + <\tau_i^{j(0)}> ,
 \end{eqnarray}
with the nonzero components
\begin{eqnarray}
\label{tik0-nut}
&&<\tau_z^{z(0)}>= -\frac{u}{24 \ell}\left(k+\frac{n^2}{\ell^2}\right)^2,~~
<\tau_\theta^{\theta(0)}>=<\tau_\varphi^{\varphi(0)}>=  
\frac{u}{12 \ell}\left(\frac{n^2}{2\ell^2}(\frac{83n^2}{\ell^2}+28 k)+k^2 \right) ,~~
\\
\nonumber
&&<\tau_\varphi^{t(0)}>= -u\frac{n}{2\ell}F_k^2(\frac{\theta}{2})\left(\frac{2n^2}{\ell^2}(\frac{36n^2}{\ell^2}+11k)+k^2\right),~~
<\tau_t^{t(0)}>= -\frac{u}{24 \ell}\left(\frac{19n^2}{\ell^2}(\frac{7n^2}{\ell^2}+2k)+k^2\right),
 \end{eqnarray}
and
\begin{eqnarray}
\label{tikf-nut}
 <\tau_z^{z(f)}>=\frac{u}{2\ell}(3c_z-c_t),~<\tau_\theta^{\theta(f)}>=<\tau_\varphi^{\varphi(f)}>=-\frac{u}{2\ell }(c_t+c_z),
 \\
 \nonumber
<\tau_\varphi^{t(f)}>=u\frac{8 c_t n}{\ell}F_k^2(\frac{m\rho}{2}),~<\tau_t^{t(f)}>=\frac{u}{2\ell}(3c_t-c_z),
\end{eqnarray}
with $u=\frac{N^2}{4 \pi^2\ell^3}$ (here we have replaced $8\pi G=4\pi^2\ell^3/N^2$
\cite{Maldacena:1997re}, with $N$ the rank of the gauge group of the
dual $\mathcal{N}=4,~d=4$ theory).
This is the form of an anisotropic perfect fluid, the first part, $<\tau_i^{j(0)}>$,
encoding the  Casimir-type background contribution.
As expected, this stress tensor is not traceless,
\begin{eqnarray}
\label{tii-nut}
  <\tau_i^i>= \frac{N^2}{4 \pi^2\ell^3}\frac{(4n^2+k\ell^2)^2}{12\ell^5},
 %<\tau_i^{i}>= \frac{(4\Omega^2+k m^2)^2}{12 m^3},
\end{eqnarray}
which
matches exactly the conformal anomaly of the boundary CFT
\cite{Skenderis:2000in}:
\beqs
{\cal A}=-\frac{2\ell^3}{16\pi G}\left(-\frac{1}{8}
{\cal R}_{ab} {\cal R}^{ab}+\frac{1}{24}{\cal R}^2\right).
\eeqs

Here one should remark that, after defining a new ``radial'' coordinate
$\rho=\ell \theta $ and the new parameters $m=1/\ell
,~\Omega =n/\ell ^{2}$,
the background metric (\ref{bound-gen}) upon which the dual field theory resides can be written as
\begin{eqnarray}  
\label{Godel-gen}
ds^{2}=d\rho^{2}+\frac{F_{k}^{2}(m\rho)}{m^{2}}d\varphi^{2}+dz^2-%
\Big(\frac{4\Omega }{m^{2}}F_{k}^{2}(\frac{m\rho}{2})d\varphi +dt%
\Big)^{2}.
\end{eqnarray}%
This can be recognised as the standard form used in the literature 
of the homogeneous G\"{o}del-type universe \cite{reboucas}.
 The famous G\"{o}del rotating
universe \cite{Godel:1949ga} corresponds to the case $k=-1,~m^{2}=2\Omega^{2},$ $(i.e.~\ell ^{2}=2n^{2})$. 
The boundary spacetime of a $k=0$ TNAdS$_5$ solution is known as the
Som-Raychaudhuri spacetime \cite{Som} (the $d=5$ counterpart of this metric is 
a supersymmetric solution \cite{Gauntlett:2002nw},
which enjoyed recently some interest in the literature).
The value $m^{2}=4\Omega ^{2}~(\ell ^{2}=4n^{2})$ with $k=-1$ is special again, and we find  
the Rebou\c{c}as-Tiomno space-time \cite{Reboucas:1986wa}, 
which is the direct product of the AdS$_3$ spacetime and the $z-$direction,
the spacetime symmetry being enhanced in this case. 

The features of the  $d=4$  G\"{o}del-type universes have been extensively discussed in the literature
from various directions. 
Of interest here is the occurrence of CTCs for a range of the parameters ($%
k,m,\Omega $) \cite{reboucas}.
The situation here reflects that found for the $d=5$ bulk spacetime.
One can easily see that for $k=1,0$ the causality is violated for any ($m,\Omega $). 
In the hyperbolic
case $k=-1$, causality violation on the boundary metric (\ref{Godel-gen})
will appear only for $m^{2}<4\Omega ^{2}$, the Rebou\c{c}as-Tiomno space-time  separating 
globally
hyperbolic spacetimes from causality violating ones.
 
 The violation of causality in a $k=0,-1$ G\"{o}del-type solution is made possible
essentially because the metric coefficient $g_{\varphi \varphi }$ assumes
negative values for $\rho>\rho_0$, (with $g_{\varphi \varphi }(\rho_0)=0$),
 an effect induced by
the nondiagonal metric term associated with rotation \cite{Hawking}. This is quite
different from creating causal anomalies in flat space by identifying the
time coordinate or from the standard AdS causal problems. 
A similar feature is shared by the Kerr black hole (for a region inside the horizon \cite{Hawking})
or by a Van Stockum infinitely long, rotating dust cylinder \cite{vanStockum:1937zz,Tipler}.
However, different from the last cases,
a  G\"{o}del-type  spacetime  (\ref{Godel-gen}) is homogeneous \cite{reboucas}, and there are CTCs through every event
(hence the causality violation is not localized to some region). Also, a
study of geodesics has shown that these spacetimes are geodesically complete
(and thus singularity free) \cite{Chandrasekhar,Calvao:1990yv}.
 
The existence of CTCs renders the formulation of a physical theory in the  backgrounds (\ref{Godel-gen})
rather obscure.
The problems in the standard quantisation of a scalar field in a G\"{o}del
 universe have been extensively discussed in
\cite{Leahy:1982dj}. The absence of a Cauchy surface results in the incompleteness 
of the mode solutions and thus in the impossibility to 
follow the standard quantisation procedure\footnote{However, 
as argued in \cite{Radu:1998sk, Astefanesei:2004kn},
it may be possible to avoid some of these problems by using the Euclidean approach to the quantum field theory. 
A Euclidean section of the line element (\ref{Godel-gen}) is found by taking
$\Omega \to i a$, $t \to i \tau$. For a general discussion of the Euclidean approach to field quantization for
acausal spacetimes, see \cite{Hawking:1995zi}, \cite{Cassidy:1997yf}.}.

However, one can use the AdS/CFT correspondence to predict
qualitative features of a quantum field theory 
in a G\"{o}del-type background, despite the fact that the quantisation procedure
is unclear in this case.
The issue  
is interesting especially in connection with the chronology
protection conjecture \cite{Hawking:1991nk}.
This conjecture is usually enforced by the back reaction of the (divergent) energy momentum
tensor of a test field on the spacetime geometry, via the semi-classical
Einstein equations.
Despite a number of attempts, the status of this conjecture in a G\"{o}del-type
background is still unclear.

As one can see from (\ref{tik-nut}),
the stress tensor of the dual theory $<\tau_i^k>$ is finite and well defined for any range of parameters.
In particular it stays finite for $\rho=\rho_0$ (where the Killing vector 
$\partial/\partial \varphi$ changes the sign for acausal $k=-1$ metrics). 
The finiteness of the stress tensor  (\ref{tik-nut}) 
implies that in this case the chronology
protection conjecture cannot be settled at this level.
This is likely to be connected  with the fact that a non-globally hyperbolic 
G\"{o}del-type
spacetime is not the result of evolution of certain initial data,
but rather it has existed "forever".
Moreover, since there is no acausal geodesic motion
in this background \cite{Chandrasekhar,Calvao:1990yv},
the standard arguments predicting a divergence of $<\tau_i^{j}>$,
based on a Hadamard form of the Green function do not apply.

A similar conclusion has been reached in \cite{Astefanesei:2004kn}
for the case of a four dimensional TNAdS metric (\ref{metric-NUT4}).
The boundary metric there is given by (\ref{Godel-gen}) with the
$dz^2$ term suppressed.
However, the structure of the boundary stress tensor 
is very different in that case from (\ref{tik-nut}) (in particular  $<\tau_i^{i}>=0$ for $d=3$).
Moreover, while the AdS$_4$/CFT$_3$ correspondence is
still relatively poor understood, this is not the case for AdS$_5$/CFT$_4$.
 
On general grounds, one  expects that a more detailed
 study of the ${\cal N}=4$ super-Yang-Mills theory
  formulated in a $d=4$ G\"{o}del-type acausal background
 would reveal the existence of some pathological features.
 For example, it would be interesting to consider the effects of the 
 higher order terms
 in the holographic stress tensor $T_{ik}$ and to look for a hydrodynamical
 description of a gauge theory in the background (\ref{Godel-gen}).
 However, these aspects would require a study beyond the framework of this paper and are
 presently under study. 

 %%%%%%%%%%%%%%%%%%%%%%%%%%%%%%%%%%%%%%%%%%%%%%%%%%%%%%%%%%%%%%%%%%
%%%%%%%%%%%%   Squashed black holes   %%%%%%%%%%%%%%%%%%%%%%%%%%%%%%
%%%%%%%%%%%%%%%%%%%%%%%%%%%%%%%%%%%%%%%%%%%%%%%%%%%%%%%%%%%%%%%%%%
\section{Squashed AdS$_5$ black holes and their generalizations}
%%%%%%%%%%%%%%%%%%%%%%%%%%%%%%%%%%%%%%%%%%%%%%%%%%%%%%%%%%%%%%%%%% 
%%%%%%%%%%%%%%%%%%%%%%%%%%%%%%%%%%%%%%%%%%%%%%%%%%%%%%%%%%%%%%%%%%
\subsection{The ansatz and asymptotics}
%%%%%%%%%%%%%%%%%%%%%%%%%%%%%%%%%%%%%%%%%%%%%%%%%%%%%%%%%%%%%%%%%%
A different class of solutions is found when  deforming the AdS$_5$ black strings  
along the $z-$direction.
The  metric ansatz in this case is
\begin{eqnarray}
\label{metric-squashed} 
ds^2= \frac{dr^2}{f(r)}+r^2(d\theta^2+F_k^2(\theta) d\varphi^2)+a(r)(dz+4n F_k^2( {\theta}/{2})d\varphi)^2-b(r)dt^2.
\end{eqnarray}
For $k=1$, these are the solutions considered in  \cite{Murata:2009jt} for a slightly different metric ansatz,
representing the natural AdS counterparts of the $\Lambda=0$ IM-type black holes \cite{Ishihara:2005dp}.
Here we clarify their asymptotics (which was not considered in \cite{Murata:2009jt}), and discuss their thermodynamics.

It is obvious that no causal pathology appears here, $t$ being a global time coordinate. For $k=1$, the coordinate
$z$ becomes essentially an Euler angle, with a periodicity
$L=8 \pi n$.
Again, for $k=0,-1$ the period of $z$ is not fixed apriori.
The range of $\theta,\varphi$ is similar to the black string case.
 
A suitable combination of the Einstein equations with a negative cosmological constant
leads to the following equations for the metric functions $a(r)$, $b(r)$ and $f(r)$:
\begin{eqnarray}
\nonumber
&&f'=\frac{2k}{r}
+\frac{8r}{\ell^2}
-\frac{2f}{r}
-f\left(\frac{a'}{a}+\frac{b'}{b} \right)-\frac{4n^2a}{r^3}~,
\\ 
\label{eqs-sq} 
&&b''=\frac{2b}{r^2} 
-\frac{2k b}{r^2f}
-\frac{4b}{ \ell^2f}
+\frac{2ba'}{ra}
+\frac{ b'}{r}
-\frac{k b'}{rf}
-\frac{4 rb'}{\ell^2f}
+\frac{a'b'}{2a}
+\frac{b'^2}{b}+\frac{2n^2a (b+r b')}{ r^4 f}~,
\\
\nonumber
&&\frac{a'}{a}= 2\frac{ b\big[ 2\ell^2(k-f)+12r^2\big]
-2 r\ell^2fb'}{r\ell^2f\big[rb'+4 b\big]}-\frac{4n^2ab}{ r^3 f(r b'+4b)}~.
\end{eqnarray}
We are mainly interested in black-hole type solutions, with a horizon located at $r=r_h>0$ 
(again, we shall not consider
the behaviour of the solutions inside the horizon). 
Near a nonextremal horizon, the following power series expansion  holds
\begin{eqnarray}
\label{exp1}
&&f(r)=f_1(r-r_h)+f_2(r-r_h)^2+O(r-r_h)^3,
~~b(r)=b_1(r-r_h)+b_2(r-r_h)^2+O(r-r_h)^3,
\\
\nonumber
&&a(r)=a_h+a_1(r-r_h)+a_2(r-r_h)^2+O(r-r_h)^3.
\end{eqnarray}
Similar to the black string case, all coefficients here are fixed by
$a(r_h)$ and $b'(r_h)$, the first terms being
\begin{eqnarray}
\nonumber
&&a_1=\frac{4a_h(2 r_h^4+a_h n^2\ell^2)}{4 r_h^5+r_h( k r_h^2-2a_h n^2)\ell^2}~,~~
f_1=\frac{4r_h}{\ell^2}+\frac{k}{r_h}-\frac{2a_h n^2}{ r_h^3},
\\
\label{exp2}
&&b_2=\frac{b_1(-8r_h^8-2r_h^4\ell^2(3k r_h^2-4a_h n^2)+(2a_h^2n^2+2k a_h n^2 r_h^2- k^2 r_h^4 )\ell^4)}
{ r_h(4 r_h^4-2a_h n^2 \ell^2+ k r_h^2\ell^2)^2},
\\
\nonumber
&&f_2=\frac{-8 r_h^8+2r_h^4\ell^2(-4a_hn^2+3k r_h^2)
+(14 a_h^2n^4-6ka_h n^2r_h^2+k^2r_h^4)\ell^4 }
{4 r_h^8 \ell^2+ r_h^4\ell^4( kr_h^2-2a_hn^2)}.
 \end{eqnarray} 
The  solutions have the following asymptotic form
\begin{eqnarray}
\nonumber
&&a(r)=\frac{r^2}{\ell^2}+\frac{1}{2}(k-\frac{4n^2}{ \ell^2}) 
+\frac{\ell^2}{r^2}\left(c_z+
\frac{1}{12}(k-\frac{4n^2}{\ell^2})(k-\frac{16n^2}{\ell^2})\log(\frac{r}{\ell})\right)+ O\left(\frac{\log r}{r^4}\right),
\\
\label{as1} 
&&b(r)=\frac{r^2}{\ell^2}+\frac{1}{2}(k-\frac{2n^2}{ \ell^2}) 
+\frac{\ell^2}{r^2}\left(c_t+
\frac{1}{12}(k-\frac{4n^2}{\ell^2})(k-\frac{8n^2}{\ell^2})\log(\frac{r}{\ell})\right)+ O\left(\frac{\log r}{r^4}\right),
\\
\nonumber
&&f(r)=\frac{r^2}{\ell^2}+\frac{2}{3} (k -\frac{5n^2}{2 \ell^2})  
+\frac{\ell^2}{r^2}\left(c_t+c_z+\frac{n^2}{2\ell^2}(k-\frac{4n^2}{\ell^2})+
\frac{1}{6}(k-\frac{4n^2}{\ell^2})(k-\frac{12n^2}{\ell^2})\log(\frac{r}{\ell})\right)+ O\left(\frac{\log r}{r^4}\right),
\end{eqnarray} 
depending again on two real parameters $c_t,c_z$.

The case $k=1,~4n^2=\ell^2$ is special, 
the Schwarzschild-AdS$_5$ solution being recovered, with
\begin{eqnarray}
\label{SAdS}
f(r)=b(r)=\frac{1}{4}+\frac{r^2}{\ell^2}+\frac{\ell^2}{r^2}c_t,~~a=\frac{r^2}{\ell^2}~.
\end{eqnarray}
In this limit, the surface of constant $r,t$ is a round sphere $S^3$.
In the general $k=1$ case, this surface is a squashed, topologically $S^3$ sphere.
However, as $n\to 0$, the topology changes to $S^2\times S^1$.

As discussed in  \cite{Murata:2009jt}, the $k=1$ configurations possess a nontrivial globally regular limit as $r_h\to 0$. 
Making the assumption of regularity at $r=0$ implies that this solution behaves near the origin as
\begin{eqnarray}
\nonumber
&&a(r)=\frac{1}{4n^2}r^2+\frac{ (1-f_2\ell^2)}{n^2\ell^2}r^4+\dots,
~~b(r)=b_0+\frac{4b_0}{\ell^2}r^2+\frac{16b_0(1-f_2\ell^2)}{3\ell^4}r^4+\dots,
\\
\nonumber
&&f(r)=\frac{1}{4}+f_2 r^2+(\frac{11f_2}{3\ell^2}-\frac{2}{3\ell^4}-3f_2^2)r^4+\dots,
\end{eqnarray}
$i.e.$ has  two free parameters $f_2$, $b_0$.
For $4n^2=\ell^2$, the usual AdS$_5$ spacetime is recovered, with $f_2=1/\ell^2$, $b_0=1/4$ in this case.

 By using the counterterm method in Section 2, we find the 
following expressions for the mass of the solutions (which contains a Casimir term $M_c^{(k)}$):  
\begin{eqnarray}
\label{M-sq} 
M&=&M_0+M_c^{(k)}~,
~~{\rm where}~~~M_0=\frac{\ell }{16\pi G 
}\big[c_z-3c_t\big]LV_{k},
~~
M_c^{(k)}= 
\frac{V_{k}L\ell}{192\pi G}(\frac{ n^2}{\ell^2}-k)^2 .
\end{eqnarray} 
The Hawking temperature and the event horizon area of these solutions are given by
\begin{eqnarray}
\label{TA} 
T_H=\frac{1}{\beta}=\frac{\sqrt{f_1b_1}}{4 \pi},~~A_H=r_h^2V_k L\sqrt{a_h}.
\end{eqnarray}  
Even in the absence 
of a closed form solution, the tree level Euclidean action $I$ of these 
solutions can be evaluated by integrating the Killing 
identity $\nabla^\mu\nabla_\nu  \zeta_\mu=R_{\nu \mu}\zeta^\mu,$
for the Killing vector $\zeta^\mu=\delta^\mu_t$, together with the  Einstein 
equation $R_t^t={(R - 2\Lambda)/2}$. In this way, one isolates the bulk action
contribution at infinity and at $r=r_h$ (or $r=0$). The divergent 
contributions given by the surface integral term at infinity are  
cancelled by the Gibbons-Hawking term in the action (\ref{action}) together with the counterterms
(\ref{ct}),
and one finds a final expression for the total action in terms of boundary data at the horizon and at infinity. 
The entropy as computed from the Gibbs-Duhem relation $S=\beta M-I$
  is $S=A_H/4G$, as expected.

%%%%%%%%%%%%%%%%%%%%%%%%%%%%%%%%%%%%%%%%%%%%%%%%%%%%%%%%%%%%%%%%%%
\subsection{Numerics and the properties of solutions}
%%%%%%%%%%%%%%%%%%%%%%%%%%%%%%%%%%%%%%%%%%%%%%%%%%%%%%%%%%%%%%%%%%
 
Again, in the absence of explicit solutions, we solved the system (\ref{eqs-sq}) numerically, as a boundary value problem. 
The methods empoyed here are similar to those described in Section 3  and we shall not enter into details. 
Again,  without any loss of generality, the AdS length scale $\ell$ is set to one and we consider only positive values of $n$.

%%%%%%%%%%%%%%%%%%%%%%%%%%%%%%%%%%%%%%%%%%%%%%%%%%%%%%%%%%%%%%%%%%%%%%%%%%%%%%%%%%%%%%%%%%%%%%
%%%%%%%%%%%%%%%%%%%%%%%%%%%%%%%%%%%%%%%%%%%%%%%
%%%%%%%%%%%%%%%%%%%%%%%%%%%%%%%%%%%%
\begin{figure}[ht]
\hbox to\linewidth{\hss%
	\resizebox{7cm}{6.1cm}{\includegraphics{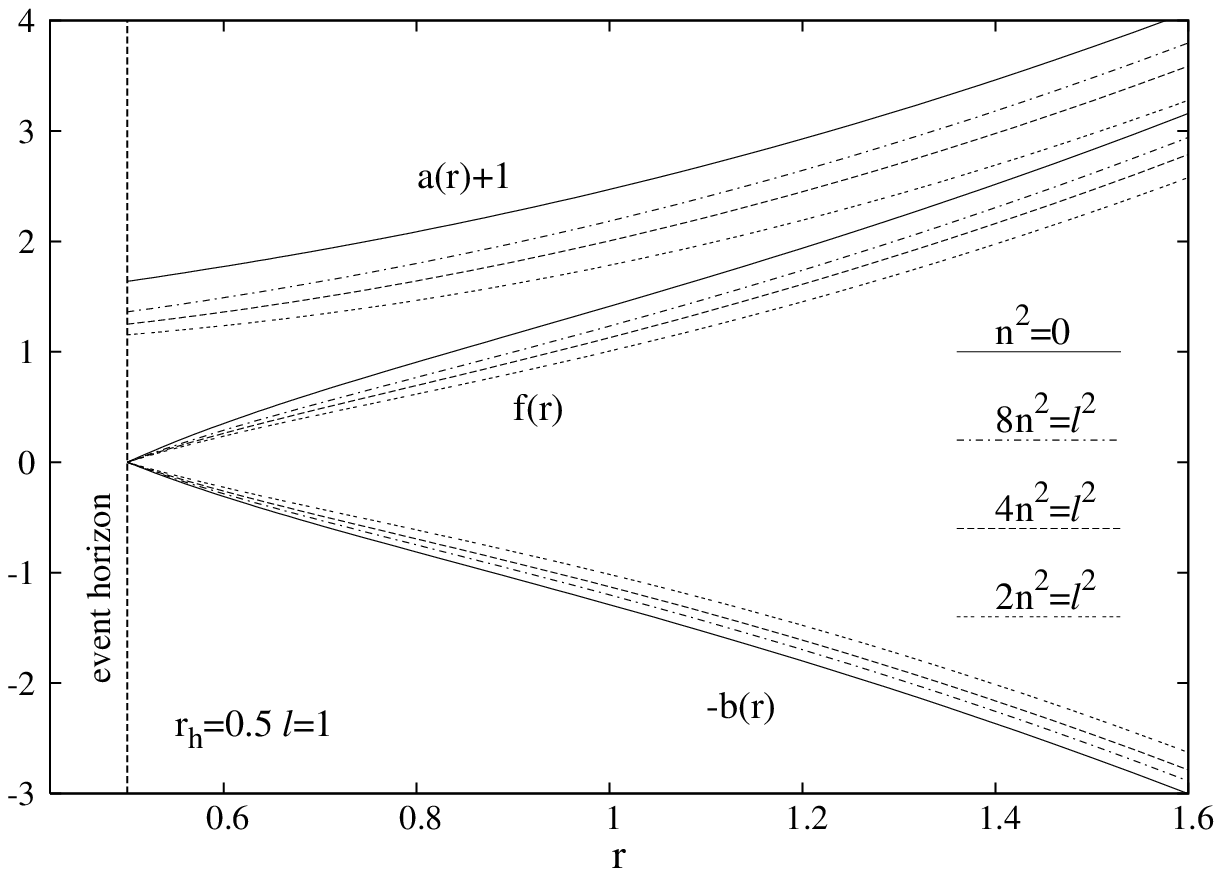}}
\hspace{5mm}%
        \resizebox{7cm}{6.1cm}{\includegraphics{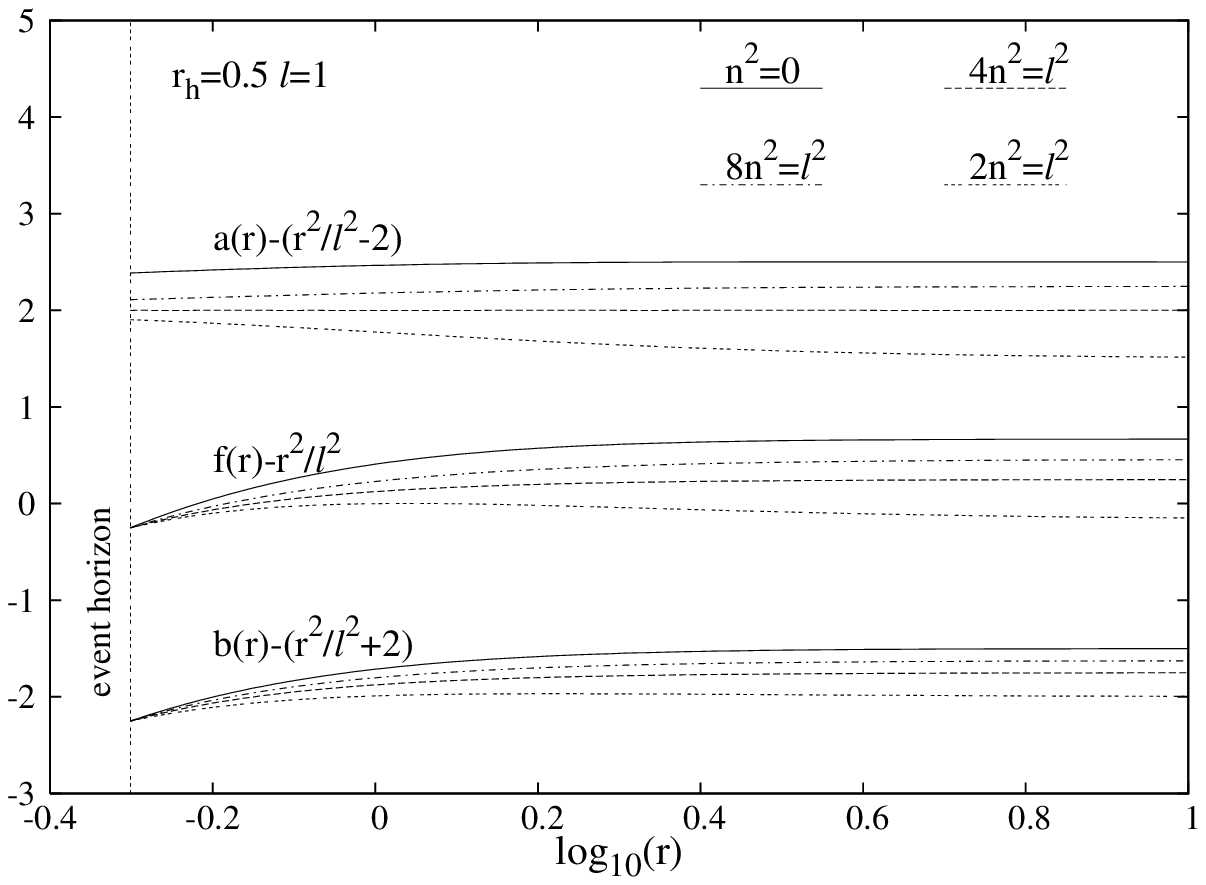}}	
\hss}
	\caption{  The profiles of typical $k=1$ black hole solutions for several values of $n$.
} 
\label{Fig1}
\end{figure}
%%%%%%%%%%%%%%%%%%%%%%%%%%%%%%%%%%%%

As in the case of nuttier configurations, our method confirms the existence, for any $k$, of families of solutions labeled 
by $r_h$ and $n$. In the limit $n \to 0$, the uniform black strings are recovered.
Again, the AdS length scale $\ell$ is set to one, without any loss of generality.

Given its potential relevance in AdS/CFT,  the case of main interest here is $k=1$.
Confirming the results in \cite{Murata:2009jt},
we could construct a family of solutions
interpolating between the AdS$_5$ uniform black strings and the Schwarzschild-AdS$_5$ black holes. 
Solutions with $n> \ell/2$ were considered as well.
Profiles corresponding to $r_h=0.5$ and several values of $n$ are presented in Figure 4 for the region close to horizon (left)
and for large values of $r$ (right).
In Figure 5 (left) we show the solutions' dependence on $n$ for $r_h=0.5$ (a similar picture was found for
several other values of the event horizon radius). 

%%%%%%%%%%%%%%%%%%%%%%%%%%%%%%%%%%%%
\begin{figure}[ht]
\hbox to\linewidth{\hss%
	\resizebox{7cm}{6.1cm}{\includegraphics{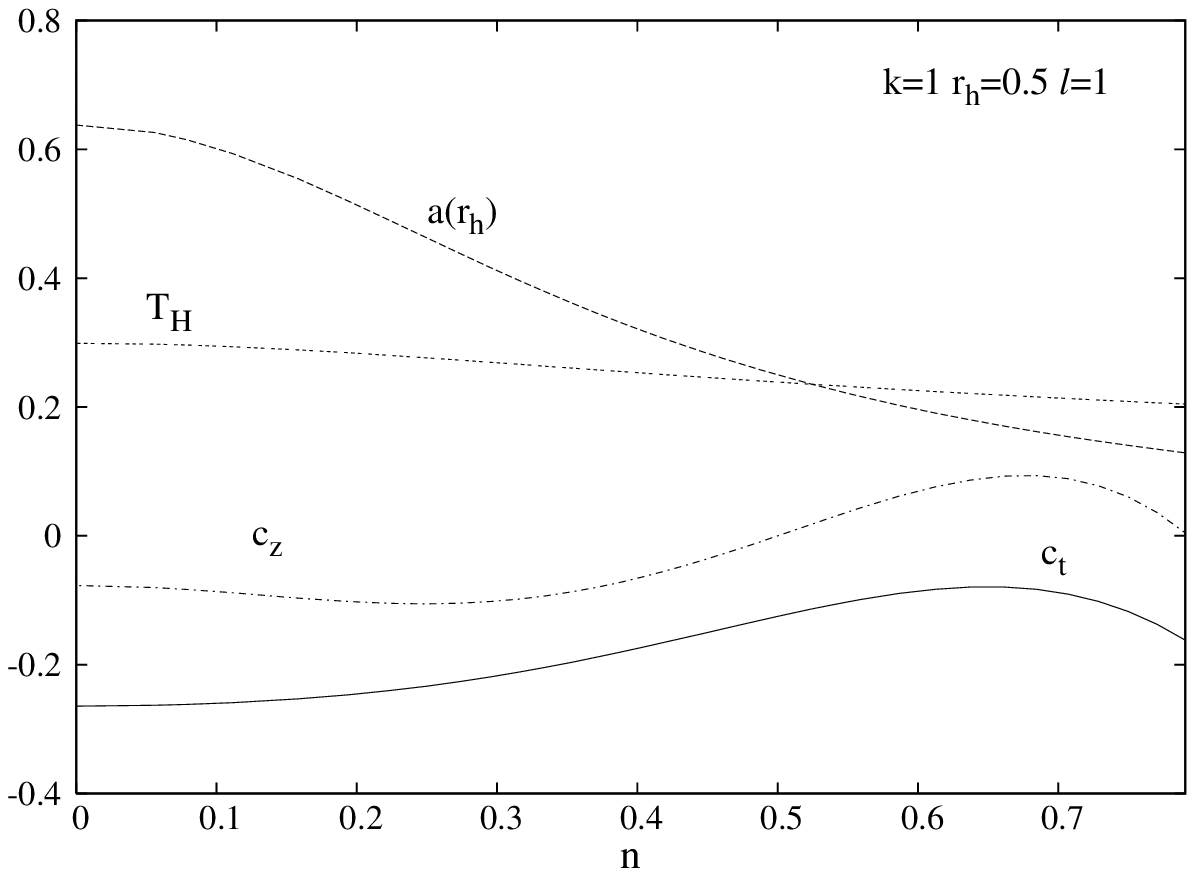}}
\hspace{5mm}%
        \resizebox{7cm}{6.1cm}{\includegraphics{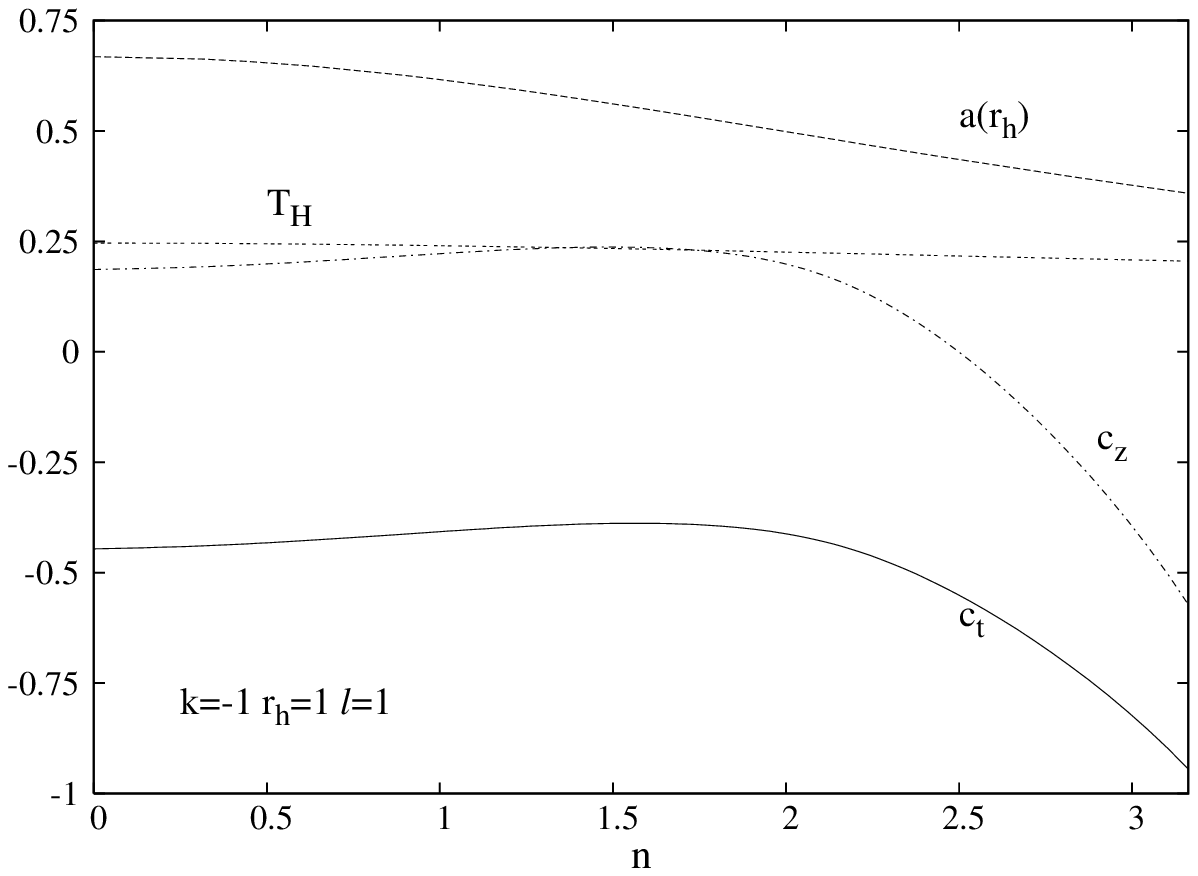}}	
\hss}
	\caption{  The dependence of the $k=1,-1$ black hole solutions on the parameter $n$ for fixed 
	event horizon radius $r_h$.
} 
\label{Fig1}
\end{figure}
%%%%%%%%%%%%%%%%%%%%%%%%%%%%%%%%%%%% 
%%%%%%%%%%%%%%%%%%%%%%%%%%%%%%%%%%%%
\begin{figure}[ht]
\hbox to\linewidth{\hss%
	\resizebox{7cm}{6.1cm}{\includegraphics{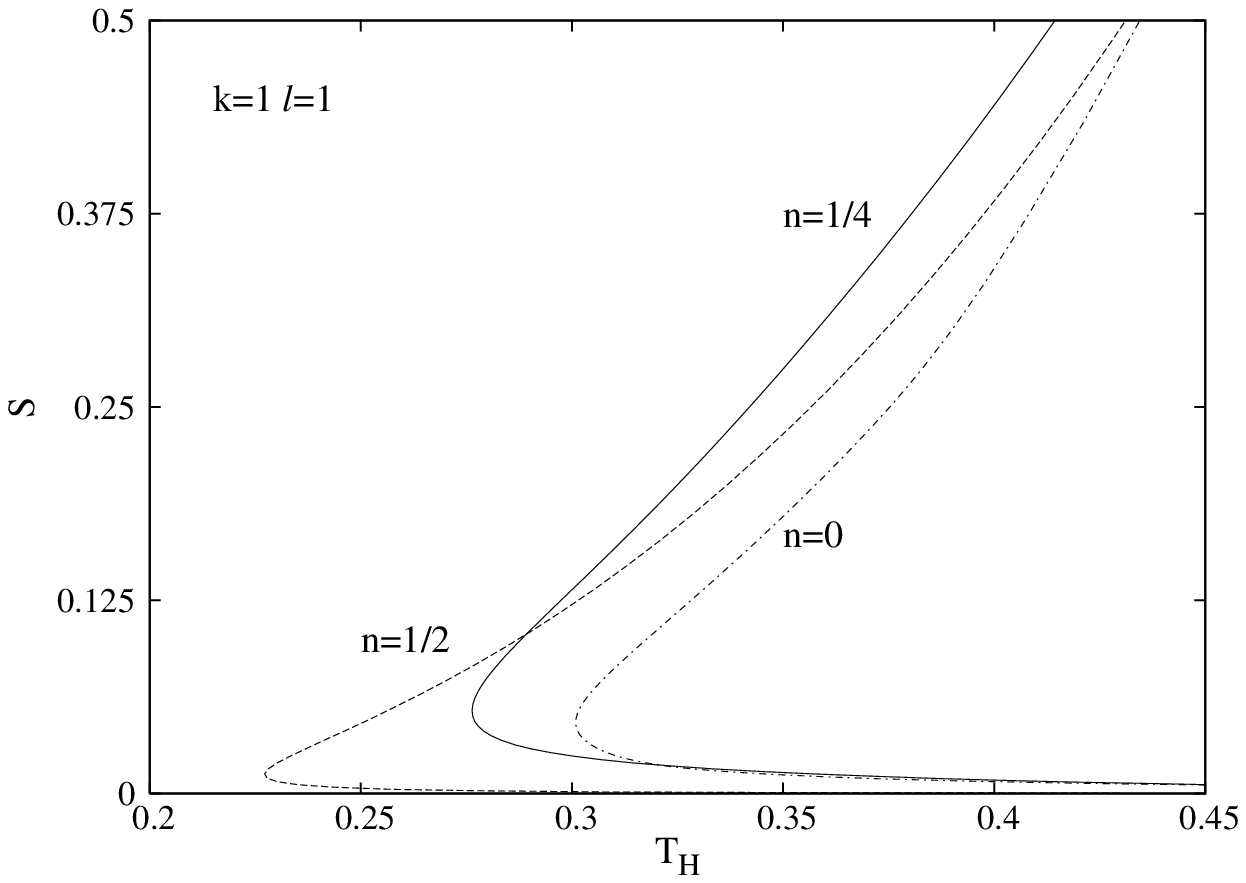}}
\hspace{5mm}%
        \resizebox{7cm}{6.1cm}{\includegraphics{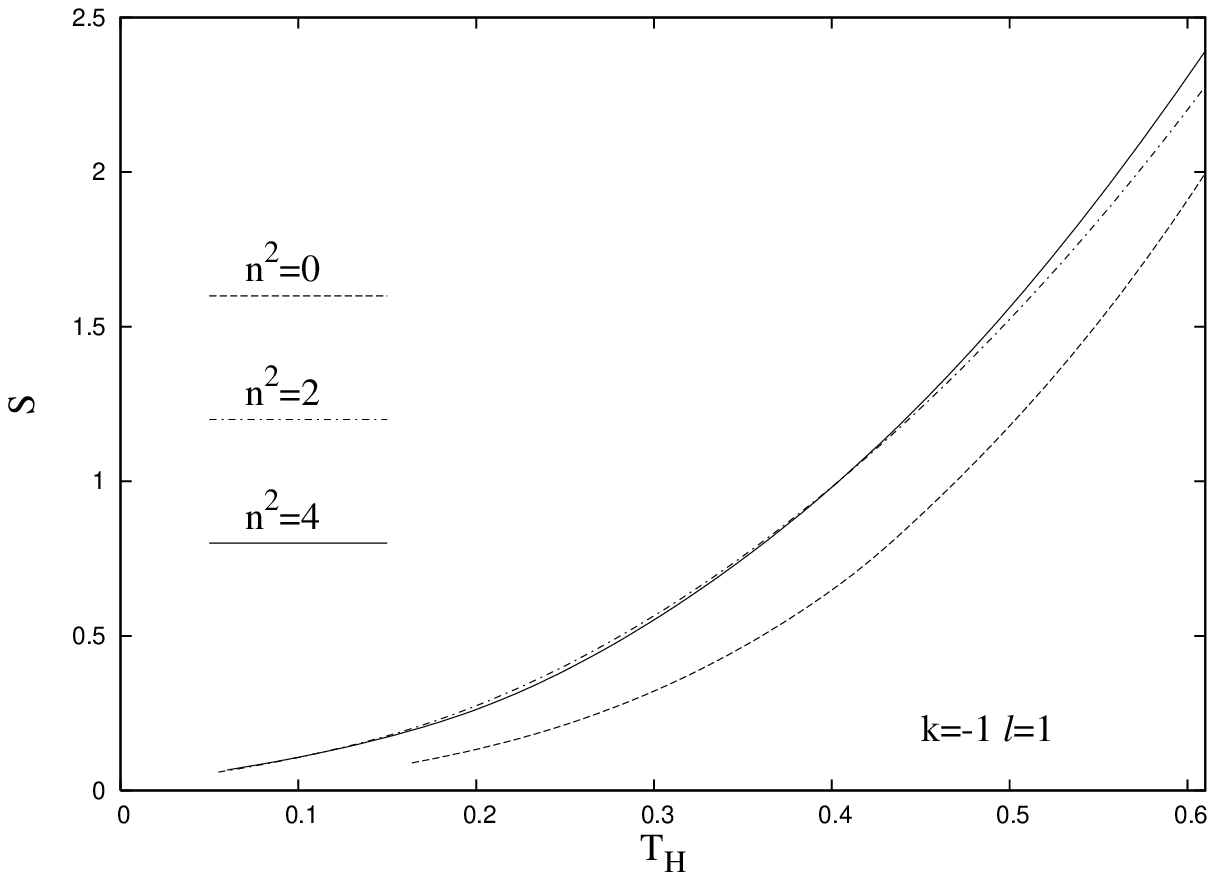}}	
\hss}
	\caption{  The entropy is plotted as a function of the Hawking temperature
	for $k=1,-1$ black holes. 
} 
\label{Fig1}
\end{figure}
%%%%%%%%%%%%%%%%%%%%%%%%%%%%%%%%%%%%
 
Concerning the thermal properties,
our numerical results indicate that
the physics familiar from the Schwarzschild-AdS$_5$ case is valid also for these $k=1$ 
squashed black holes.
Their temperature  is bounded from below for any $n$, and 
 we have two 
branches consisting of smaller (unstable) and large (stable) black 
holes (see Figure 6 (left)). 
Thus, at low temperatures we have 
a single bulk solution, which corresponds to the 
thermal globally regular solution (whose temperature may take arbitrary values).
The free energy $F=I/\beta$ of the $k=1$ solutions is positive for small $r_h$ 
and negative for large $r_h$. This shows that the phase transition 
found in \cite{Hawking:1982dh} occurs also in this case, 
as already conjectured in \cite{Murata:2009jt}.  
This is illustrated in Figure $7$, where the free energy is plotted versus the temperature for several values of $n$.
 
 %%%%%%%%%%%%%%%%%%%%%%%%%%%%%%%%%%%%
\begin{figure}[ht]
\hbox to\linewidth{\hss%
	\resizebox{9cm}{7.1cm}{\includegraphics{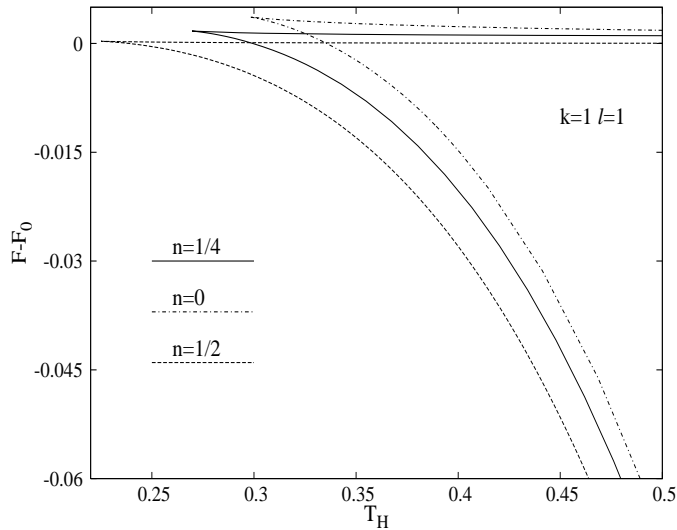}}
% \hspace{5mm}%
%        \resizebox{7cm}{6.1cm}{\includegraphics{sq-k-1-var-rh.eps}}	
 \hss}
	\caption{   The free energy \textit{vs.} the temperature for the small 
and large $k=1$ squashed black hole solutions is plotted for three values of $n$. Here we have subtracted the free 
energy contribution $F_0$ of the corresponding globally regular solutions.
} 
\label{Fig1}
\end{figure}
 We have also considered solutions with $k=-1,0$, corresponding to
topological black holes. These types of configurations
have a number of common features with their $k=1$ counterparts.
For example, the typical shape of the metric functions is similar to those presented in Figure 4.
However, for any value of $n$,
the $k=-1,0$ solutions exist only for large enough values of the horizon radius $r_h$.
Their thermal properties are also different, as shown for the  $k=-1$ solutions in Figure 6 (right).
In this case we noticed the existence of only one branch of thermally stable black holes, $i.e.$
their entropy increases with the temperature.
This is the behaviour found in  \cite{Mann:2006yi} for the $n=0$ topological UBS limit 
(this feature is  also familiar from the study of the usual Schwarzschild-AdS topological black holes).

The dependence on the parameter $n$ of a number of global quantities is plotted  in Figure 5 (right) for 
$k=-1$ solutions with $r_h=1$.

 %%%%%%%% 

%%%%%%%%%%%%%%%%%%%%%%%%%%%%%%%%%%%%%%%%%%%%%%%%%%%%%%%%%%%%%%%%%%
\subsection{The boundary metric and the dual CFT}
%%%%%%%%%%%%%%%%%%%%%%%%%%%%%%%%%%%%%%%%%%%%%%%%%%%%%%%%%%%%%%%%%% 
In the asymptotic region, the $k=1$ metric (\ref{metric-squashed})
becomes
\begin{eqnarray}
\label{asympt-metric-squashed1}
ds^2=\frac{\ell^2}{r^2}dr^2+r^2\left( d\theta^2+\sin^2 \theta d \varphi^2
+\frac{4n^2}{\ell^2}(d\psi+\cos \theta d\varphi)^2-\frac{1}{\ell^2}dt^2 \right),
 \end{eqnarray}
 where $\psi=\varphi-z/(2n)$.
 One can recognize  that a surface of constant ($r,t$)
 represents a squashed three sphere, with $4n^2/\ell^2$ parametrising the squashing.
 The value $n=2\ell$ separates prolate metrics from the oblate case ($n>2\ell$).
  Remarkably, this asymptotic metric is still maximally symmetric, $i.e.$
 to leading order
 $R_{ijkl}=-1/\ell^2(g_{ik}g_{jl}-g_{il}g_{jk})$ 
 (this holds also for $k=0,-1$ metrics in the large $r$ limit).
 
The  general form of
the four dimensional boundary metric is given by the line element  
 \begin{eqnarray}
\label{bm-squashed1}
ds^2= \ell^2(d\theta^2+F_k^2(\theta) d\varphi^2) + (dz+4n F_k^2( {\theta}/{2})d\varphi)^2- dt^2,
 \end{eqnarray}
 which is a homogeneous spacetime with five Killing vectors, the $k=1$ case corresponding to the frozen mixmaster universe.
  
Some features of the boundary CFT for $k=1$ squashed black hole metrics were already discussed in  \cite{Murata:2009jt}.
Here we give the expression of the  boundary stress tensor for dual CFT
in a background metric (\ref{bm-squashed1}): 
\begin{eqnarray}
\label{tik}
<\tau_a^{b}>= <\tau_a^{b(f)}>+<\tau_a^{b(0)}>,
 \end{eqnarray}
with the nonvanishing 
components\footnote{Not by accident, 
this form reminds of the stress tensor (\ref{tik-nut})-(\ref{tikf-nut}) in a  G\"{o}del-type background.
The reason is that line-elements (\ref{bound-gen}), (\ref{bm-squashed1}) present the same
Euclidean section and are related through a simple
analytic continuation involving also the parameter $n$.
Thus the expressions of $<\tau_a^{b}>$ are also related. 
 A similar situation 
 appears $e.g.$ for the case of Misner spacetime and the infinitely long cosmic string solution \cite{Cassidy:1997yf}.}
\begin{eqnarray}
\nonumber
&&<\tau_z^{z(0)}>= -\frac{u}{24\ell} \left( \frac{19 n^2}{\ell^2}(\frac{7 n^2}{\ell^2}-2k)+k^2 \right),~~
<\tau_\theta^{\theta(0)}>= <\tau_\varphi^{\varphi(0)}>= 
\frac{u}{12\ell}\left( \frac{ n^2}{2\ell^2}(\frac{83n^2}{\ell^2}-28k)+k^2 \right),~~
\\
\label{tik0}
&&<\tau_\varphi^{z(0)}>= -u\frac{ n }{2\ell}(\frac{18n^2}{\ell^2}- k)(\frac{4n^2}{\ell^2}- k)F_k^2(\theta/2),
<\tau_t^{t(0)}>=-\frac{u}{24\ell}\left( \frac{ n^2}{\ell^2}- k \right)^2,
 \end{eqnarray}
and
\begin{eqnarray}
\label{tikf}
 <\tau_z^{z(f)}>= u\frac{3c_z-c_t}{2\ell},~ 
<\tau_\theta^{\theta(f)}>=<\tau_\varphi^{\varphi(f)}>= -u\frac{c_t+c_z }{2\ell} ,
\\
\nonumber
<\tau_\varphi^{z(f)}>= u\frac{8 c_z n }{\ell }F_k^2(\frac{\theta}{2}),
~
<\tau_t^{t(f)}>= u\frac{3c_t-c_z}{2\ell},
 \end{eqnarray}
 where, again,  $u= {N^2}/{4 \pi^2\ell^3}$.
One can easily verify that the trace of this tensor  
%\begin{eqnarray}
%\label{tii-nut}
%$ <\tau_i^i>= \frac{N^2}{4 \pi^2\ell^3} {(k\ell^2-4n^2)^2}/({12\ell^5}),$ 
%\end{eqnarray}
matches exactly the conformal anomaly of the boundary CFT.

As discussed in \cite{Murata:2009jt}, the thermodynamical entropy
of the ${\cal N}=4$ super-Yang-Mills theory formulated in a $k=1$ background (\ref{bm-squashed1})
agrees with that of the $k=1$ squashed black holes up to a factor of $3/4$.
It would be interesting to extend the results in  \cite{Murata:2009jt} to the cases $k=0,-1$.
This computation looks possible, since the line-element (\ref{bm-squashed1})
is homogeneous for any $k$, which allows an explicit computation of the spectrum
for scalar, gauge or spinor fields.
A computation of the $ <\tau_i^k>$ tensor 
for the ${\cal N}=4$ super-Yang-Mills theory is another reasonable task,
at least for $k=1$,
given the existence in the literature
of a number of partial results in this case \cite{Critchley:1981am}.

%%%%%%%%%%%%%%%%%%%%%%%%%%%%%%%%%%%%%%%%%%%%%%%%%%%%%%%%%%%%%%%%%%
\section{Further remarks}
%%%%%%%%%%%%%%%%%%%%%%%%%%%%%%%%%%%%%%%%%%%%%%%%%%%%%%%%%%%%%%%%%%
In this paper we have presented arguments 
for the existence of two different generalizations of
the known AdS$_5$  black string solutions.
The first type of configuration, discussed in Section 3, can be interpreted as
the uplifted version of  four dimensional
TNAdS solutions and inherits most of the properties in that case.
In particular, only a restricted set of solutions with a hyperbolic base space
are free of causal pathologies.

An interesting question here is the possibility of the existence
of more general nuttier solutions with a dependence on
the coordinate $z$. 
These configurations would be intrinsic five-dimensional, and 
would present new qualitative features.
Indeed, such configurations are known to exist for $k=1$
in the $n=0$ limit, and describe static nonuniform AdS
black strings \cite{Delsate:2008kw,Delsate:2008iv}.
This issue is related to 
the classical stability of the uniform solutions against small perturbations.
 The results in \cite{Brihaye:2007ju} indicate that the small AdS black strings whose 
conformal infinity is the product of time and $S^{d-3}\times S^1$
  possess a classical  Gregory-Laflamme instability \cite{Gregory:1993vy}.
Their nuttier generalizations are likely to exhibit this
instability as well, at least for some range of the parameters. 
 This would imply also the existence of a complicated phase structure of the solutions
 for the same set of boundary conditions, with a new branch of $d=5$ 
nuttier  nonuniform solutions,   possessing
 a bulk dependence on both $r$ and $z$. 
 
In Section 4, we discussed the basic properties of another type of solutions
representing
$d=5$ squashed black holes and their topological generalizations.
The main result there is a proof that the phase transition 
found in \cite{Hawking:1982dh} for AdS$_5$ black holes with spherical horizon
 occurs also for solutions with a squashed horizon.  

Our solutions in Sections 3 and 4 can be used as new test grounds for the AdS/CFT correspondence.
They have a nontrivial boundary structure which is generically a circle fibration 
over a base space that can have exotic topologies.
Therefore their study could shed some light on the study of CFTs on some unusual backgrounds.
In particular, one could be able to understand the thermodynamic phase structure of 
CFTs by working out the corresponding 
phase structure
of the solutions in the bulk.
An unexpected result here is the emergence of the $d=4$ G\"odel spacetime 
as the boundary metric of a particular nuttier solution.
This may lead to some progress in the issue of field quantization in the
presence of causal pathologies.

We close this work with several remarks on possible extensions of the solutions in this paper.
For example, in Section 3 we have considered the Lorentzian form of the nuttier configurations.
However, the main relevance of the NUT charged solutions is for a 
Euclidean signature of spacetime, which would allow a  study of 
their thermodynamics.
This may be an interesting subject, since 
the results in \cite{Astefanesei:2004kn} suggest that the
entropy/area relation is always violated
in the presence of a NUT charge.
In principle, the Euclidean section of the nuttier solutions 
is simply obtained using the analytic continuations $%
t\rightarrow i \chi$ and $n \rightarrow i n$.
However, in the absence of a general exact solution, this
is not useful in practice\footnote{This can easily be seen looking
$e.g.$ at the $d=4$ closed form solution (\ref{metric-NUT4}), (\ref{mf-NUT4}).
The   metric functions $f(r)$, $b(r)$ look  different on the Lorentzian and Euclidean sections.},
and one has to employ again numerical methods to study the  Euclidean counterparts of the solutions in
Section 3.
One should also remark that, on the Lorentzian section, the mass and NUT parameter are unrelated.  
That is, once we fix the event horizon radius, the parameter $n$ can
be freely specified. 
However, the $k=1$ Euclidean solutions have to satisfy one extra regularity condition,
since no conical singularity should appear both at $r=r_h$ and $\theta=\pi$. This implies the relation
$8\pi p n=4\pi/\sqrt{f'(r_h)b'(r_h)}$, with $p$ an integer \cite{Mann:2003zh}.
From a numerical point of view,  this would restrict the allowed values of $r_h$ for a given $n$, and thus the
possible values of $M$.
Moreover, the possible existence of both "nut" and "bolt" configurations for a Euclidean signature
is likely to lead to a complicated landscape of solutions there.

The analytic continuation  $z \to i \tau$, $t\rightarrow i \chi$ and $n \rightarrow i n$ of a
 nut-charged solution  in Section 3
 leads to another interesting type of configuration, with 
a line element
\begin{eqnarray}
\label{bubble-NUT} 
ds^2= \frac{dr^2}{f(r)}+r^2(d\theta^2+F_k^2(\theta) d\varphi^2)
+b(r)\Big(4 n F_{k}^{2}(\frac{\theta}{2})d\varphi+d\chi\Big)^{2}-a(r)d\tau^2, 
\end{eqnarray}
(we recall that the functions $f(r), b(r)$ vanish  for some $r=r_h>0$, while $a(r_h)>0$; thus these configurations
are different from the squashed black holes in Section 4).
These solutions can be interpreted as 
 nuttier deformation of the AdS bubbles (\ref{bubble}).
 For $k=1$ and $n=\ell/2$, an exact solution here reads
 \begin{eqnarray}
f(r)=\frac{r^2}{\ell^2}+\frac{1}{4}+\frac{\ell^2}{r^2}c_t+\frac{\ell^4}{4r^4}c_t,~~~
b(r)=\frac{r^2}{\ell^2}+ \frac{ \ell^2}{r^2}c_t,~~~~a(r)=\frac{r^2}{\ell^2}+\frac{1}{4},
 \end{eqnarray}
 which is the AdS$_5$ soliton discussed in \cite{Clarkson:2006zk}.
The above configuration if free of singularities (apart from the central one at $r=0$)
 for $c_t=-(4p^2-1)^2/(256 p^4)$, with $p$ an arbitrary integer.
 As argued in \cite{Clarkson:2006zk}, this solution is the 
 lowest energy state within that asymptotic class.
 The results in this work suggest the existence of similar AdS$_5$ solitons
 for any value of the ratio $n/\ell$.
 We hope to return elsewhere to this subject.

Moreover, one expects other types of nut charged solutions to exist. For example,
in principle one can use the globally hyperbolic solutions in Section 4 
to generate new  nuttier solutions by using the analytic 
continuation $t\to i t$, $n \to i n$, $z \to iz$
for the line element (\ref{metric-squashed}).  
The asymptotic structure of the resulting configurations would  be similar to that of the solutions in Section 3.
However, they would satisfy a different set of boundary conditions than (\ref{eh-nut}) at $r=r_h$,
with $a(r_h)=0$ and $b(r_h)>0$. 
Again, any progress in this direction appears to require a separate numerical study of the solutions.

%%%%%%%%%%%%%%%%%%%%%%%%%%%%%%%%%%%%%%%%%%%%%%%%%%%%%%

\section*{Acknowledgements}

Y. B. thanks  the Belgian FNRS for financial support. 
The work of ER was supported by a fellowship from the Alexander von Humboldt Foundation. 
%%%%%%%%%%%%%%%%%%%%%%%%%%%%%%%%%%%%%%%%%%%%%%%%%%%%%%%%%%%%%%%%%%%%%%%%%%
 
%%%%%%%%%%%%%%%%%%%%%%%%%%%%%%%%%%%%%%%%%%%%%%%%%%%%%%%%%%%%%%%%%%%%%%%%%%%%%

\end{document}